\documentclass[conference, two column]{IEEEtran}
\usepackage{cite}
\usepackage{amsmath,amssymb,amsfonts}
 
\usepackage{algorithmic}
\usepackage{graphicx}
\usepackage{textcomp}
\usepackage{comment}
\usepackage{color}
\usepackage{epstopdf}
\usepackage{listings}
\usepackage{amsthm}

\usepackage{float}
\usepackage{color} 
\definecolor{mygreen}{RGB}{28,172,0} 
\definecolor{mylilas}{RGB}{170,55,241}
\usepackage{float} 

\usepackage[center]{caption} 
\usepackage{multicol}
\usepackage{mathtools} 
\usepackage{cuted} 
\usepackage[ruled,vlined]{algorithm2e}
\usepackage{hyperref}

\usepackage{lipsum}
\usepackage{enumitem}
\usepackage{nomencl}
\makenomenclature

\begin{document}
	\title{Maximum Correntropy Polynomial Chaos Kalman Filter for Underwater Navigation}
\author{
\IEEEauthorblockN{Rohit Kumar Singh\IEEEauthorrefmark{1}, Joydeb Saha \IEEEauthorrefmark{2}, Shovan Bhaumik \IEEEauthorrefmark{3}}\\
\IEEEauthorblockA{\IEEEauthorrefmark{1}\IEEEauthorrefmark{2}\IEEEauthorrefmark{3}Department of Electrical Engineering, Indian Institute of Technology Patna, India}
\IEEEauthorblockA{\IEEEauthorrefmark{1} \emph{rohit\_1921ee19@iitp.ac.in,}\IEEEauthorrefmark{2} \emph{joydeb\_2121ee32@iitp.ac.in,}\IEEEauthorrefmark{3} \emph{shovan.bhaumik@iitp.ac.in}}

}

\maketitle
\thispagestyle{plain}
\pagestyle{plain}
\begin{abstract}
This paper develops an underwater navigation solution that utilizes a strapdown inertial navigation system (SINS) and fuses a set of auxiliary sensors such as an acoustic positioning system, Doppler velocity log, depth meter, attitude meter, and magnetometer to accurately estimate an underwater vessel's position and orientation. The conventional integrated navigation system assumes Gaussian measurement noise, while in reality, the noises are non-Gaussian, particularly contaminated by heavy-tailed impulsive noises. To address this issue, and to fuse the system model with the acquired sensor measurements efficiently, we develop a square root polynomial chaos Kalman filter based on maximum correntropy criteria. The filter is initialized using acoustic beaconing to accurately locate the initial position of the vehicle. The computational complexity of the proposed filter is calculated in terms of flops count. The proposed method is compared with the existing maximum correntropy sigma point filters in terms of estimation accuracy and computational complexity. The simulation results demonstrate an improved accuracy compared to the conventional deterministic sample point filters.
 \end{abstract}

\nomenclature[01]{\(b\), \(g\), \(i\), \(e\), \(n\)}{Body frame, geodetic frame, inertial frame, ECEF and NED/navigation frame of reference, respectively.}
\nomenclature[02]{\(P^{g}=[  L \ l \ Z  ]\)}{Position vector \emph{viz.} latitude ($L$), longitude ($l$) and depth ($Z$) in geodetic frame.}
\nomenclature[03]{\(V^{n}=[v^{N} \ v^{E} \ v^{D}]\)}{Velocity vector in local NED frame.}
\nomenclature[04]{\(\Psi^b=[\phi \; \theta\; \psi]\)}{Attitude in body frame, \emph{\emph{i.e}.} roll ($\phi$), pitch ($\theta$) and yaw ($\psi$).}
\nomenclature[05]{\(\omega^{b}=\dot{\Psi}^b\)}{Body rate in the body frame.}
\nomenclature[06]{\(C_{b}^{n}\)}{Rotation matrix from body to navigation frame.}
\nomenclature[07]{\(\omega_{ie}^n\)}{Angular velocity of the inertial frame \emph{wrt} ECEF expressed in the navigation frame.}
\nomenclature[08]{\(f^{b}\)}{Force vector in the body frame.}
\nomenclature[09]{\(g^n\)}{Acceleration due to gravity in NED frame.}
\nomenclature[10]{\(R_M\)}{Meridian radius of curvature at a latitude $L$.}
\nomenclature[11]{\(R_N\)}{Transverse radius of curvature at a latitude $L$.}
\nomenclature[12]{\(R_E\)}{Length of semi-major axis of the earth.}
\nomenclature[13]{\(\epsilon\)}{Eccentricity of the ellipsoid of the earth.}
\nomenclature[14]{\(y_k\)}{Measurement vector at $k^{th}$ instant.}
\nomenclature[15]{\(\mu_k\), \(r_k\)}{Process and measurement noise, respectively.}
\nomenclature[16]{\(\hat{X}_{k\vert k-1}\), \(\hat{X}_{k\vert k}\)}{Prior and posterior state estimate.}
\nomenclature[17]{\(P_{k\vert k-1}\), \(P_{k \vert k}\)}{Prior and posterior error covariance matrix.}
\nomenclature[18]{\(P_{yy,k\vert k-1}\), \(P_{Xy,k\vert k-1}\)}{Prior innovation covariance matrix and Prior cross-covariance matrix.}
\nomenclature[19]{\(K_k\)}{Kalman gain.}
\nomenclature[20]{\(\kappa (.)\), \(\sigma\)}{Gaussian kernel function and kernel bandwidth.}
\nomenclature[21]{\(\mathcal{E}_k\)}{Weighted error matrix.}
\nomenclature[22]{\(\bar{P}_{k\vert k-1}\)}{Modified prior error covariance matrix.}
\nomenclature[23]{\(\bar{R}_k\)}{Modified measurement noise covariance matrix.}
\nomenclature[24]{\(\mathcal{X}_j\)}{Set of sample points.}
\nomenclature[25]{\(\xi\)}{Matrix containing all the collocation points.}
\nomenclature[26]{\(H_p\)}{Hermite polynomial of order $p$.}
\nomenclature[27]{\(\hat{A}\), \(\hat{B}\)}{Coefficient matrix for state and measurement function.}

\printnomenclature[1.2in]

\begin{section}{Introduction}
Due to the unavailability of satellite-based positioning systems, limited visibility and the absence of continuous communication with surface-based system, underwater navigation remains a substantially challenging task \cite{zhuang2023multi}. In aerospace, substantial advancements have been made in the realm of navigation. However, a noticeable disparity becomes evident when one shifts their focus to the field of underwater navigation. This discrepancy can be attributed to the inherent complexity and challenges associated with underwater navigation due to which the literature addressing this remains rather limited. The review papers \cite{kumar2021recent,zhang2023autonomous} discussed about various target tracking algorithms based on active and passive modes of operation and compared them. \cite{sahoo2019advancements}, \cite{paull2013auv} present different autonomous underwater vehicles (AUV) navigation methods and their challenges. 

Navigation with dead reckoning system \cite{leonard2016autonomous} which integrates the output of an inertial navigation system (INS) to calculate the location and orientation of a vehicle is erroneous and the estimates diverge due to sensor noise, biases and uncertainties in initialization \cite{brokloff1997dead,groves2015principles}.
 In literature, fusion of INS is attempted with terrain-based navigation \cite{kinsey2006survey}, magnetism-based navigation \cite{zhang2020geomagnetic}, gravity-based navigation \cite{wang2016particle}, and visual matching \cite{duecker2020towards}.
However, there are several shortcomings of these methods such as the limited availability of suitable data sets and dynamic behaviour of physical features of deep sea environment \cite{sobreira2019map}. To mitigate such disadvantages, INS data are generally fused with onboard auxiliary sensors \cite{paull2013auv}, \cite{miller2010autonomous,davari2016asynchronous}.  

In this paper, a methodology is proposed to fuse the INS data and dynamic model of an underwater vehicle with Doppler velocity log (DVL), acoustic positioning system (APS), pressure sensor, attitude sensor and magnetometer using nonlinear estimators. A six degree of freedom model of vehicle is adopted from the earlier literature \cite{allotta2016unscented}. However, unlike the earlier literature \cite{davari2016asynchronous,allotta2016unscented}, we assume no GPS signal is available. We use  APS along with other onboard sensors to initialise the estimators. The APS system which relies on the transmission and reception of acoustic signals between the transponders and receivers, provides the location of an underwater vehicle as long as the vehicle lies in the the operational range of it \cite{zhang2023autonomous}.

In all the previous works, it is assumed that the sensor noises follow a Gaussian distribution, which is not always true in real life \cite{urooj20222d}. Here, we consider sensor noises as non-Gaussian and characterize them with a weighted sum of Gaussians. As the noises become non-Gaussian, the traditional Gaussian estimators such as the extended Kalman filter (EKF), unscented Kalman filter (UKF) \cite{wan2000unscented}, cubature Kalman filter (CKF) \cite{arasaratnam2009cubature}, new sigma point Kalman filter (NSKF) \cite{radhakrishnan2018new}, Gauss-Hermite filter (GHF) \cite{bhaumik2019nonlinear}, polynomial chaos Kalman filter (PCKF) \cite{kumar2023polynomial} do not provide adequate accuracy while estimating the position, velocity and orientation of an underwater vehicle.  Although the particle filter can handle non-Gaussian noises, but its computational load is very high making it unsuitable for onboard application. Here, we propose a maximum correntropy based non-linear filter which uses a set of deterministic points (collocation point), generated from polynomial chaos expansion and it is named as maximum correntropy polynomial chaos Kalman filter (MC-PCKF).

We know that the system model of an underwater navigation problem is non-linear, and here the sensor noises are non-Gaussian.  In such a situation our proposed MC-PCKF has ability to outperform the existing maximum correntropy criterion based non-linear filters such as MCC-UKF \cite{liu2016maximum}, MCC-CKF \cite{liu2018maximum} and MCC-NSKF \cite{urooj20222d}; because the nonlinearity is approximated by polynomial chaos expansion (PCE)\cite{xu2018novel} in an efficient way and the maximum correntropy criteria \cite{chen2017maximum} takes care the non-Gaussian noise.  During the execution of MC-PCKF, Cholesky decomposition is required to calculate and during which a round-off error is introduced due to digital computer's limited word length capacity. In certain cases, due to such rounding, Cholesky decomposition can lead to negative definiteness matrix \cite{van2001square}, consequently terminating the filtering process. To overcome this, a square-root version of MC-PCKF is designed, which avoids Cholesky decomposition by propagating the square root of the covariance matrix \cite{arasaratnam2009cubature},\cite{bhaumik2019nonlinear}.

The performance of square root MC-PCKF is compared to existing maximum correntropy filters like MC-UKF, MC-CKF, and MC-NSKF for a simulated trajectory. It is observed that the proposed filter performs better than the existing filters in terms of root mean square error (RMSE) and averaged RMSE value. It is worth mentioning here that the filters take APS measurement data at the beginning of estimation and the measurements are unavailable as the vehicle moves away from it. When APS signal is unavailable, the system becomes unobservable, as a result, errors of position estimate start slowly increasing. Moreover, the computational burden of the proposed filter is comparable to other maximum correntropy based deterministic sample point filters.

So, we see that it is extremely important to know a vehicle's own location and orientation at any point in time. In an underwater environment, that task becomes difficult due to (i) nonlinear process dynamics, (ii) unavailability of GPS signal, and (iii) non-Gaussian sensor noises. Under such challenges, an efficient algorithm is warranted which will provide reasonably good results, and this paper is introduced for such a purpose. Hence, the objectives of this paper are to design an efficient filtering algorithm for fusing sensor data with vehicle dynamics in GPS-denied environment without revealing the own vehicle location to enemy ship or submarine. In view of the above objectives, the contribution made in this paper is summarized as follows: (i) a real-life underwater navigation problem is formulated, (ii) a new filter entitled MC-PCKF and its square root version are proposed to estimate the states in presence of a heavy-tailed non-Gaussian measurement noise, (iii) an initialization technique using APS is incorporated during filters implementation, (iv) the proposed filter shows improvement in performances in underwater navigation problem as compared to all existing filters.
\end{section}
 \section{Problem Formulation} \label{probemformulation} 
\subsection{Process Model}	
The strapdown inertial navigation system (SINS) relies on inertial measurement unit (IMU), which is equipped with accelerometers and gyroscopes to measure the vehicle's linear accelerations and body rates. The measurements from IMU are not accurate and are generally corrupted with sensor noises \cite{mu2021practical}. These data are then integrated through the kinematic model which consists of a set of differential equations to obtain the vehicle's position and attitude and they are given as:
\begin{equation} \label{eq2} 
			\dot{X} = f(X,u) + w,
\end{equation}
 where $X=[P^g \ V^n \ \Psi]^\top$ is the state vector of the system and  $f(.)$ is the process function. $u = [f^b \ \omega_{ib}^b]^\top$ is the input vector and $w$ is the process noise due to uncertainty in the control inputs, which is assumed as white, Gaussian with zero mean and $Q$ covariance \emph{\emph{i.e}.} $w \sim \mathcal{N}(0,Q)$. It is worthy to note that the input vector $u$ mentioned in Eqn. \eqref{eq2}, consists of input acceleration $f^b$ along all the three axes. The input accelerations are the resultant acceleration which acts on the body and it includes the hydrodynamic forces such as buoyancy, drag \emph{etc}.
 The output of the IMU sensor includes the true value of measurements along with Gaussian white noise, as given below. The output of the accelerometers $\tilde{f}^{b}$ and gyroscopes $\tilde{\omega}_{ib}^b$ in the body frame are given as
\begin{equation} \begin{split}
			\tilde{f}^{b} & = f^b+w_a,\\
			\tilde{\omega}_{ib}^b & = \omega_{ib}^b+w_g,
	\end{split} \end{equation}
where $w_a$ and $w_b$ are Gaussian with zero mean and covariance $\sigma_{a}^2$ and $\sigma_g^2$, respectively. The biases in the accelerometer and drift in the gyroscope are neglected. It is to be noted that the mathematical notations superscripts are used to identify the frame of reference in which vectors are represented. For example, $\omega_{xy}^z$ is read as the angular rate vector in frame $y$ with respect to frame $x$ as represented in frame $z$.

The position and the velocity follow the following differential equations: 
 \begin{equation} \label{eq3}
		\begin{split}
			\begin{bmatrix}  \dot{L} \\ \dot{l} \\ \dot{Z} \end{bmatrix} & = \begin{bmatrix} \frac{1}{R_N+Z} & 0 & 0 \\
				0 & \frac{1}{\cos L(R_E+Z)} & 0 \\ 0 & 0 & -1 \end{bmatrix} \begin{bmatrix}
				v^{N} \\ v^{E} \\ v^{D}
			\end{bmatrix},
		\end{split}
	\end{equation} 
 and 
  \begin{equation} \label{eq4}
		\begin{bmatrix}  \dot{v}^N \\ \dot{v}^E \\ \dot{v}^D \end{bmatrix}  =  C_b^n \begin{bmatrix} f_x^b \\ f_y^b \\ f_z^b \end{bmatrix} + \begin{bmatrix} 0 \\ 0 \\ g^n  \end{bmatrix} - ( 2\omega_{ie}^n+\omega_{en}^n )  \begin{bmatrix} v^N \\ v^E \\ v^D  \end{bmatrix},
	\end{equation}
 where  $g^n$ is the acceleration due to gravity in the NED frame. Interesting to note that the force vector $f^b$ is measured by the accelerometer of the IMU sensor in the body frame. So to get the velocity in NED frame it is required to transform it in the same frame. This is done by using  the direction cosine matrix  \cite{groves2015principles}, $C_b^n$, which is further expressed as 
 	\begin{equation} \label{eq6} 
		C_b^n= \begin{bmatrix} c_\theta c_\psi & -c_\phi s_\psi+s_\phi s_\theta c_\psi & s_\phi s_\psi+c_\phi s_\theta c_\psi \\
			c_\theta s_\psi & c_\phi c_\psi+s_\phi s_\theta s_\psi & -s_\phi c_\psi+c_\phi s_\theta s_\psi \\
			-s_\theta & s_\phi c_\theta & c_\phi s_\theta
	\end{bmatrix}, \end{equation} 
where $c_\theta$ and $s_\theta$ represent $\cos\theta$ and $\sin\theta$ respectively. 
 The last term of (\ref{eq4}) arises due to the Coriolis effect, which is incorporated by a skew-symmetric matrix $(2\omega_{ie}^n+\omega_{en}^n )$ whose expressions are given as $\omega_{ie}^n =  \begin{bmatrix}
			\boldsymbol{\omega} \cos L & 0 & -\boldsymbol{\omega} \sin L
		\end{bmatrix},$ and $\omega_{en}^n =  \begin{bmatrix} \frac{v^E}{R_N+Z} & -\frac{v^N}{R_M+Z} & -\frac{v^E\tan L}{R_N+Z}
		\end{bmatrix}.$
The Earth's rotational velocity is $\boldsymbol{\omega} = 7.2921150 \times 10^{-5} \text{rad/sec}$ in the ECEF frame, which is directed along the Earth's rotational axis towards the north pole. 

The orientation vector follows the following differential equation: 
\begin{equation}\label{eq5} \begin{split}
		\begin{bmatrix}
			\dot{\phi} & \dot{\theta} & \dot{\psi}
		\end{bmatrix} ^\top & = \Omega^{-1} \begin{bmatrix}
			\omega_x & \omega_y & \omega_z
		\end{bmatrix}^\top\\
      & = \Omega^{-1} ( \omega_{ib}^b-C_b^n [ \omega_{ie}^n+\omega_{en}^n ]  ),
	\end{split} 
 \end{equation}	 
where $\Omega$ in (\ref{eq5}) is the Earth rotation matrix, given as 
\begin{equation*}
		\Omega = \begin{bmatrix}
			1 & 0 & -\sin \theta \\ 0 & \cos\phi & \sin\phi \cos\theta \\ 0 & -\sin\phi & \cos\phi\cos\theta
		\end{bmatrix}.
\end{equation*}
The continuous time INS equation from (\ref{eq2}) is given as
 \begin{equation}  \label{eqdiscrete} \begin{split}
     \begin{bmatrix}
         \dot{P}^g \\ \dot{V}^n \\ \dot{\Psi}^b
      \end{bmatrix} & = \begin{bmatrix}
          0_{3 \times 3} & diag(\frac{1}{R_M+Z}, \frac{1}{\cos L(R_N+Z)} , -1) & 0_{3 \times 3} \\
          0_{3 \times 3} & - ( 2\omega_{ie}^n+\omega_{en}^n )_{\times} & 0_{3 \times 3} \\
          0_{3 \times 3} & 0_{3 \times 3} & 0_{3 \times 3} \end{bmatrix} 
         \\ &  \begin{bmatrix} P^g \\ V^n \\ \Psi^b \end{bmatrix}  +\begin{bmatrix}
              0_{1 \times 3} & C_b^nf^b+g^n & \Omega^{-1}\omega_{ib}^b \end{bmatrix}^\top +w.
  \end{split} \end{equation}	
It is well known that the DCM becomes singular near $\frac{\pi}{2}$ rad and it leads to a numerical instability. This is known as the gimbal lock \cite{hemingway2018perspectives}. For this reason the quaternion representation of attitude is preferred particularly for an aerospace problem \cite{crassidis2006sigma}. However, in this work, we adopt the DCM representation due to its simplicity, ease of implementation and we ensure that due to vehicle's limited body movement, the gimbal lock never arises. 
\subsection{Measurement Model}
In this work, we use measurements of velocity obtained from Doppler velocity log \cite{farrell2008aided}, measurements of depth obtained from depth sensor \cite{klein2015observability}, roll and pitch manipulated from accelerometer measurements, and measurements of yaw from magnetometer sensor \cite{healey1998online}. To see where in the vessel, the sensors are generally mounted, interested readers may see \cite[pp. 434]{farrell2008aided}. All the measurements are corrupted with measurement noise which are assumed as additive non-Gaussian. As in our process model state vector is in NED frame, if the measurements are not received in NED frame they are converted to NED frame of reference by multiplying them with an appropriate transformation matrix \cite{farrell2008aided,shaukat2021multi}.   

The velocity measurements obtained from DVL are in body frame of reference. So they are converted to NED frame and the velocity measurement equation becomes  
	\begin{equation} \label{eq7}
		y_{1,k} = V^n_k = C_b^n V_{m,k}^b+r_{v,k},
	\end{equation}
where $V_{k}^b$ is the velocity measured in the body frame at $k^{th}$ instant, $r_{v,k}= [r_{v^N} \ r_{v^E}  \  r_{v^D}]^\top$ is the measurement noise which is to be modelled as a weighted sum of Gaussian noises.
	
The depth is measured by a pressure sensor and is given by 
	\begin{equation}\label{eq8}
		y_{2,k} = Z_k+ r_{z,k},
	\end{equation}
where $r_{z,k}$ is noise in depth measurement, which is a weighted sum of Gaussian noises.
	
During constant velocity motion, the acceleration due to gravity is the only acceleration acting on the body. In such a case, the roll angle ($\phi_{k}$) and pitch angle ($\theta_{k}$) are obtained using the $\theta_{k} = \frac{\sin^{-1}(f_{x,k}^b)}{g^n}$ and $\phi_{k} = -\frac{\sin^{-1}(f_{y,k}^b)}{g^n\cos\theta_k}$ respectively, where $f_{x,k}^b$ and $f_{y,k}^b$ are the acceleration acting on the body along the forward and starboard direction.
The measurement of roll and pitch angles are given as
\begin{equation} 
 	y_{3,k} = \begin{bmatrix} \phi_k \\ \theta_k \end{bmatrix} +\begin{bmatrix} r_{\phi,k} \\ r_{\theta,k} \end{bmatrix} ,
\end{equation}
where $r_{\phi,k}$ and $r_{\theta,k}$ are sensor noise which is to be modelled as the weighted sum of Gaussian noises. 

The yaw measurement is given by:
	\begin{equation} \label{eqn9}
		y_{4,k} =  \psi_k +r_{\psi,k}.
	\end{equation}
Due to magnetic disturbances from sources such as undersea currents and surrounding metal objects, magnetometer measurements are subjected to errors, which combinedly are represented with  $r_{\psi,k}$. Further, it is assumed as a weighted sum of Gaussian noises. 

\subsubsection{Measurement model I}
Combining the Eqns. \eqref{eq7}-\eqref{eqn9}, the measurement model becomes 
 \begin{equation} \label{mm1}
      y_k  =  \begin{bmatrix} y_{1,k} & y_{2,k} & y_{3,k} & y_{4,k} \end{bmatrix}^\top  
      =  h(X_k)+r_k, 
  \end{equation}
where $r_k = \begin{bmatrix}
    r_{v,k} & r_{z,k} & r_{\phi,k} & r_{\theta,k} & r_{\psi,k}  \end{bmatrix}^\top$ is the non-Gaussian measurement noise vector. It is interesting to note that the rank of the observability Gramian for the system defined by (\ref{eq2}) to (\ref{mm1}) (after linearization) is 7. So the system position states, \emph{i.e}. latitude and longitude are unobservable and estimation error for such states are unlikely to converge. 
 \subsubsection{Measurement model II} 
 \begin{figure}[ht]  
\centerline{\includegraphics[scale =0.55]{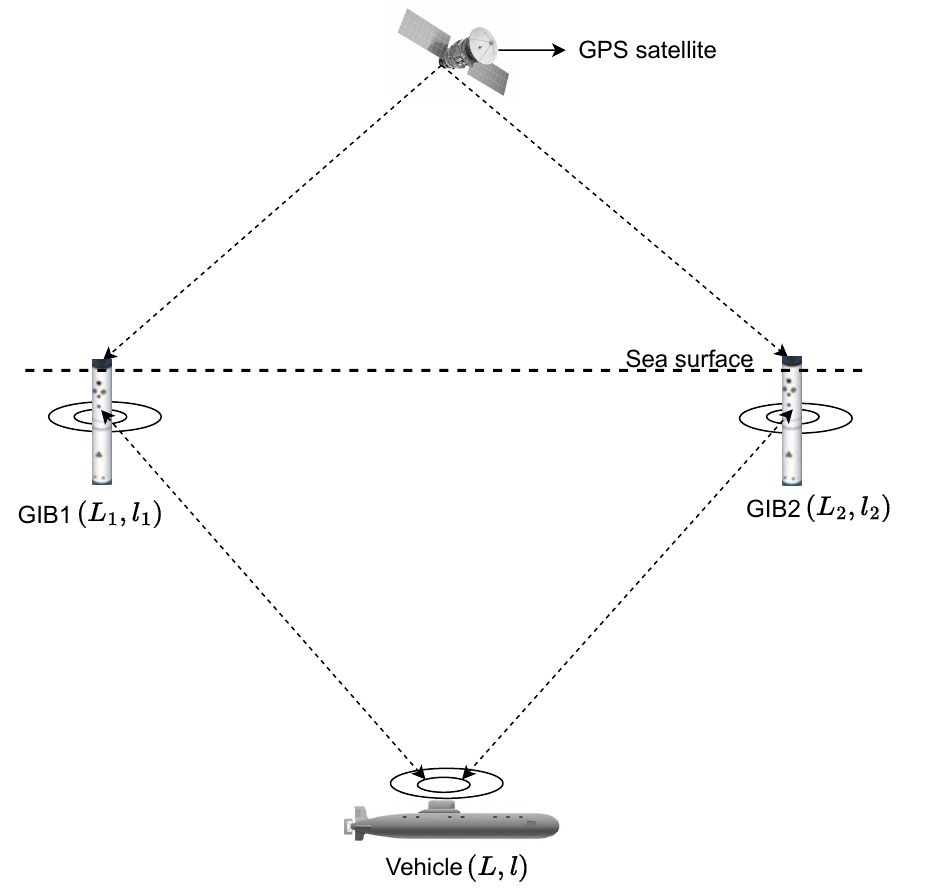}}
\caption{Acoustic positioning system for underwater navigation.}
\label{underwater1} \end{figure} 
As we see the unavailability of position measurements makes the system unobservable. GPS sensors revolutionized navigation on land/sea surface and in the air in terms of their accuracy in localization. The inability of GPS signals to penetrate water, combined with signal attenuation, multipath interference, and the lack of line-of-sight, renders GPS impractical for underwater navigation. Their limitations in underwater environments have led to the adoption of acoustic positioning systems \cite{marco2001command}. 

Acoustic positioning system (APS) uses sound waves that propagate well underwater and can penetrate through the water, providing reliable positioning information even in challenging underwater environments. Stationary buoys or beacons on the sea surface transmit signals to underwater vehicle, allowing the vehicle to locate itself \cite{zhang2023autonomous}. 
To make the system observable so as to estimate the position accurately, we make use of an APS measurement sensor, along with the set of sensors discussed in measurement model I.
 APS provides the latitude and longitude measurements, but due to its limited operating range capability, it can only be primarily used for the initialization of the estimators \cite{kebkal2017auv}. Measurements from the APS are available to the estimators only up to a certain time step as long as the vehicle lies within the transmitting and receiving range of the acoustic signals from the GPS intelligent buoys (GIBs) \cite{kinsey2007situ}. The schematic representation of APS is shown in Fig. \ref{underwater1}. The measurement equation modelling for the APS is briefed below.

The values of latitude and longitude of the underwater vehicle are obtained using the triangulation method in a geodetic frame of reference \cite{paull2013auv}. Suppose the acoustic beacons, GIB1 and GIB2, have the fixed location at $(L_{1},l_{1})$ and $(L_{2},l_{2})$, whose location is known with some uncertainty. The measurement model to get the latitude and longitude of underwater vehicle is:
	\begin{equation} \label{eqAPS}
		y_{5,k} = \begin{bmatrix}  \frac{L_2\beta_{1,k}-L_1\beta_{2,k}+(l_1-l_2)\beta_{1,k}\beta_{2,k}}{\beta_{1,k}-\beta_{2,k}} \\ \frac{L_2-L_1+l_1\beta_{1,k}-l_2\beta_{2,k}}{\beta_{1,k}-\beta_{2.k}} 
		\end{bmatrix} + \begin{bmatrix}
		    r_{L,k} \\ r_{l,k} \end{bmatrix},
	\end{equation} 
where $[ r_{L,k} \ r_{l,k} ]^\top$ are non-Gaussian noises used to incorporate the sensor noise and uncertainty in the acoustic beacons location. $\beta_{1,k}$ and $\beta_{2,k}$ are the bearings of underwater vehicle obtained by beaconing from two transponders, GIB1 and GIB2, located at the sea surface, as shown in Fig. \ref{underwater1}. The bearing is obtained as $\beta_{1,k} = \tan^{-1} ( \frac{x_{k}^N-x^N_1}{x^E_k-x^E_1} )$ and $\beta_{2,k} = \tan^{-1} ( \frac{x^N_k-x^{N}_2}{x^{E}_k-x^{E}_2})$, where $(x_{k}^N,x_{k}^E,x_k^D)$ is the local NED equivalent to the geodetic coordinate. $(x_{1}^N,x_{1}^E,x_1^D)$ and $(x_{2}^N,x_{2}^E,x_2^D)$ is the location of GIB1 GIB2, respectively. The conversion of position from geodetic to NED frame is mentioned below:
\begin{equation}
    P^n = C_e^n (P^e-P^e_{ref}),
\end{equation}
   where $P^e$ is the equivalent of the geodetic coordinate in the ECEF frame given as \cite{farrell2008aided}:
   \begin{equation*}
       P^e = \begin{bmatrix}
           (R_E+Z) \cos{L} \cos{l} \\ (R_E+Z) \sin{L} \cos{l} \\ [R_E(1-\epsilon ^2)+Z]\sin{l}
       \end{bmatrix}.
   \end{equation*}

  Combining  \eqref{mm1} and \eqref{eqAPS}, the measurement model can be written as
  \begin{equation} \label{eqmm2} \begin{split}
      y_k & = h(X_k)+r_k\\
      & = \begin{bmatrix} y_{1,k} & y_{2,k} & y_{3,k} & y_{4,k} & y_{5,k} \end{bmatrix}^\top + r_k,
  \end{split}\end{equation}
where $r_k = \begin{bmatrix}
    r_{v,k} & r_{z,k} & r_{\phi,k} & r_{\theta,k} & r_{\psi,k} & r_{L,k} & r_{l,k} \end{bmatrix}^\top$ is the measurement noise vector that follows a non-Gaussian distribution modelled as weighted sum of Gaussian noises.
     The observability Gramian matrix for the non-linear process model and measurement model described in equations (\ref{eq2}) and (\ref{eqmm2}) is of full rank. It is observed that all the system states remain fully observable as long as the position measurements from the APS are available.
     
The integrated underwater navigation system estimates the underwater vehicle's position, velocity and orientation through sensor fusion with the measurement models I and II, using the Kalman filtering approach. In this work, we are implementing the nonlinear filters where the estimators use the vehicle's position and orientation as a state vector to predict and update them using sensor measurements \cite{miller2010autonomous},\cite{lee2007simulation}. The accuracy of the estimated state depends on the quality of the sensor measurements, the accuracy of the dynamic model, and the efficiency of the state estimators. The efficiency of the state estimation algorithm is particularly important as the system is nonlinear, and there is no optimal estimator for a nonlinear system that can tackle the presence of non-Gaussian noises in the sensor data. An estimator is designed in the next section, considering these points.
	
\subsection{Correntropy Function}
Correntropy is a similarity measure between two probability distributions. For two random variables X and Y with joint probability density function $f_{XY}(x,y)$, the correntropy is given by
	\begin{equation} \begin{split}
			\mathcal{V}(X,Y) & = \mathbb{E}[\kappa(X,Y)] \\
			&= \int_x\int_y \kappa(x,y)f_{XY}(x,y)dxdy,
	\end{split}\end{equation}
where $\mathbb{E}$ is the expectation, and $\kappa (.)$ is a kernel function \cite{chen2017maximum}. The kernel function takes two data points as input and it returns a scalar value that represents the similarity between them. The kernel function is of user choice and here we have considered a Gaussian kernel function given by
\begin{equation} \label{eqkernel}
\kappa(x,y) = G_{\sigma}(e)= \exp (-\frac{(x-y)^2}{2 \sigma^2}), 
\end{equation}
where $e=x-y$, and $\sigma$ represents the kernel bandwidth. In practical scenario, instead of joint pdf $f_{XY}$, a limited number of data $[(x_i,y_i)]_{i=1}^N$ are available. In such cases, we use a sample mean estimator to write the correntropy for each variable and we take its summation over the samples to define the total correntropy,  
\begin{equation}
\mathcal{\hat{V}}(X,Y) = \frac{1}{N} \sum_{i=1}^{N}G_\sigma(e_i).
\end{equation}
Taking the Taylor series expansion of kernel function, we get, 
\begin{equation}
		\mathcal{V}(X,Y) = \frac{1}{\sqrt{2\pi}\sigma} \sum_{n=0}^{\infty} \frac{(-1)^n}{2^nn!} \mathbb{E}[ \frac{(X-Y)^{2n}}{\sigma^{2n}}].
\end{equation}
As it can be observed from the above expression, the correntropy contains the weighted sum of even ordered moments of $(X-Y)$ and $\sigma$ is the weighting parameter. 
Although, both the mean square error (MSE) and correntropy are used to quantifying the dissimilarity between two probability distributions, MSE only accounts for the second-order moment and correntropy, on the other hand, contains the higher order moments of the error function. Due to consideration of higher order moments, correntropy can accommodate impulsive or non-Gaussian noise \cite{liu2007correntropy}. Further, as $\sigma$ increases, impact of higher order terms reduces significantly, and the second order moment will become the only significant term, and correntropy will approach the MSE.

\subsection{General framework for maximum correntropy filtering} \label{MCKF}
From the discussion in the previous Section, we see that our underwater navigation problem is a nonlinear estimation problem where both the process and measurement model become nonlinear. In a general way in discrete time we can write them as 
\begin{equation} \label{eq17}\begin{split}
	X_k &= f(X_{k-1})+\mu_k, \\
	y_k &= h(X_k)+r_k,
	\end{split}\end{equation} 
where $X_k\in \mathbb{R}^{n}$ and $y_k\in \mathbb{R}^{m}$ denote the state and measurement vectors respectively. The process noise, $\mu_k$ is Gaussian, with zero mean and covariance $Q_k$ \emph{viz.}  $\mu_k\sim \mathcal{N}(0, Q_k)$. The measurement noise, $r_k$ follows a non-Gaussian distribution that is modelled as a sum of weighted Gaussian, with equivalent covariance $R_k$. The estimation of states for a system defined above can be performed in two steps: (i) time update, and (ii) measurement update. 
\subsubsection{time update}
The prediction step uses the state transition model
to project the current state estimate forward in time. The predicted pdf or the prior pdf of states is obtained by Chapman-Kolmogorov equation \cite{bhaumik2019nonlinear},
\begin{equation} \label{bayes1}
    p(X_k|y_{1:k-1})= \int p(X_k|X_{k-1}) p(X_{k-1}|y_{k-1})dX_{k-1}.
\end{equation}
The prior estimate is the mean of the prior pdf and it is expressed as
\begin{equation}\label{eqnprior}  \begin{split}
    \hat{X}_{k|k-1} = & \mathbb{E}[X_{k}|y_{1:k-1}]
     = \mathbb{E}[f(X_{k-1})|y_{1:k-1}]\\
    = & \int f(X_{k-1})p(X_{k-1}|y_{1:k-1})dX_{k-1}.
\end{split} \end{equation}
Although, due to nonlinearity, the posterior pdf of previous step no longer remains Gaussian, in this work, we assume it as Gaussian. Under such assumption, the prior mean  and prior error covariance can be written as  
\begin{equation}\label{eqx}
   \hat{X}_{k|k-1} = \int f(X_{k-1})\mathcal{N}(X_{k-1};\hat{X}_{k-1|k-1},P_{k-1|k-1})dX_{k-1}, 
\end{equation}
and 
\begin{equation}\label{eqpk}\begin{split}
    P_{k|k-1} = & \mathbb{E}[(X_k-\hat{X}_{k|k-1})(X_k-\hat{X}_{k|k-1})^\top|y_{1:k-1}]\\
    = & \int f(X_{k-1})f^\top(X_{k-1})\mathcal{N}(X_{k-1};\hat{X}_{k-1|k-1}\\ 
    & P_{k-1|k-1})dX_{k-1}-\hat{X}_{k|k-1}\hat{X}_{k|k-1}^\top +Q_k.
\end{split}\end{equation}

\subsubsection{Measurement update}
 To accommodate a non-Gaussian measurement noise in the filtering process, a maximum correntropy cost function is proposed. At first, we augment the non-linear system model with measurement \cite{chen2017maximum}, as below: 
\begin{equation} \label{eq21}
		\begin{bmatrix} 
			\hat{X}_{k|k-1} \\ y_k	\end{bmatrix} = \begin{bmatrix} X_k \\ h(X_k)
		\end{bmatrix} + \nu_k,
\end{equation}
where $\nu_k = \begin{bmatrix}  -(X(k) -\hat{X}_{k|k-1}) \\ r_k \end{bmatrix}$. 
Now, 
\begin{equation} \label{eq22n} \begin{split} 
	\mathbb{E}(\nu_k\nu_k^\top) & = \begin{bmatrix} P_{k|k-1} & 0 \\ 0 & R_k \end{bmatrix} = \begin{bmatrix}  \mathcal{S}_{k|k-1}\mathcal{S}_{k|k-1}^\top & 0 \\ 0 & 	\mathcal{S}_{R,k}\mathcal{S}_{R,k}^\top \end{bmatrix},\\
& =  \mathcal{B}_k\mathcal{B}_k^\top,
\end{split} 
\end{equation} 
where $\mathcal{B}_k = diag(\mathcal{S}_{k|k-1} , \mathcal{S}_{R,k})$, $\mathcal{S}_{k|k-1}$ and $\mathcal{S}_{R,k}$ are square-roots of $P_{k|k-1}$ and $R_k$, respectively. 
Left multiplying both side of equation (\ref{eq21}) by $\mathcal{B}_k^{-1}$, the recursive model in equation (\ref{eq21}) is transformed to a non-linear regression model given as
\begin{equation} 
 \mathcal{B}_k^{-1} 	\begin{bmatrix} \hat{X}_{k|k-1} \\ y_k	\end{bmatrix} =  \mathcal{B}_k^{-1} \begin{bmatrix} X_k \\ h({X}_{k})	\end{bmatrix} +  \mathcal{E}_k, 
 \end{equation} 
 where $\mathcal{E}_k$ is the weighted error matrix given by
   \begin{equation} \label{eq23}
			 \mathcal{E}_k =  \mathcal{B}_k^{-1} \nu_k = \begin{bmatrix}
				-\mathcal{S}_{k|k-1}^{-1} (X_k-\hat{X}_{k|k-1}) \\ \mathcal{S}_{R,k}^{-1} (y_k-h({X}_{k}))  \end{bmatrix}.
	\end{equation}

We consider a maximum correntropy based cost function as follows: 
\begin{equation} \label{eq24}
		\mathclap{J}_{MCC}(X_k)  = \frac{1}{n+m} \sum_{j=1}^{n+m} exp ( \frac{-e_{k,l}^2}{2\sigma^2}),
\end{equation}
where $e_{k,l}$ represents the $l^{th}$ element of weighted error matrix, $\mathcal{E}_{k}$. From (\ref{eq23}), we say that $e_{k,l}$ represents the weighted state prediction error for $l=1:n$ and weighted measurement error from $l=n+1:n+m$, such that
\begin{equation}\label{eqekj}\begin{split}
     e_{k,l} = & -\mathcal{S}_{k|k-1}^{-1} (X_k-\hat{X}_{k|k-1}), \ l= 1,2\ldots n,\\
     e_{k,l} = & \mathcal{S}_{R,k}^{-1} (y_k-h({X}_{k})), \ l=n+1\ldots n+m.
 \end{split}     
 \end{equation}
The posterior state estimate $\hat{X}_{k|k}$ is obtained by maximizing the cost function (\ref{eq24}) \emph{i.e}.
\begin{equation} \label{eq25}
		\hat{X}_{k|k} = \arg \; \max_{X_k} \ J_{MCC}(X_k).
\end{equation}
 Now, we use the present measurement and define a weighting parameter called correntropy matrix, $\mathit{\Pi}_k$, as \cite{liu2018maximum,chen2017maximum,liu2016extended,liu2016maximum,liu2019linear,liu2007correntropy,singh2010closed,zhao2022robust}
  \begin{equation}
      \mathit{\Pi}_k = \begin{bmatrix}
          \Pi_{P,k} & 0 \\ 0 & \Pi_{R,k}
      \end{bmatrix}, \end{equation}
 where $\mathit{\Pi}_{k,j}=\exp(-\frac{e_{k,l}^2}{2\sigma^2})$ and 
      \begin{equation} \label{eq29} \begin{split}
			\Pi_{P,k} & = diag (\begin{bmatrix} \mathit{\Pi}_{k,1} & \mathit{\Pi}_{k,2} & . . & \mathit{\Pi}_{k,n} \end{bmatrix}_{1\times n}),\\
			\Pi_{R,k} & = diag (\begin{bmatrix} \mathit{\Pi}_{k,n+1} & \mathit{\Pi}_{k,n+2} & . . & \mathit{\Pi}_{k,n+m} \end{bmatrix}_{1\times m}).
	\end{split} 
 \end{equation}
  So, the modified prior error covariance and modified measurement noise covariance are 
\begin{equation} \label{eqp} 
    \bar{P}_{k|k-1} = \mathcal{S}_{k|k-1}\Pi_{P,k}^{-1}\mathcal{S}_{k|k-1}^\top,
 \end{equation}
 and 
 \begin{equation} \label{eqr} 
    \bar{R}_{k} = \mathcal{S}_{R,k}\Pi_{R,k}^{-1}\mathcal{S}_{R,k}^\top,
\end{equation}
 respectively. The calculation of correntropy matrix,  $\mathit{\Pi}_k$ is an extra step required for maximum correntropy filtering and for this square roots of $P_{k|k-1}$, and $R_k$ are to be calculated. 
 
Now we use Bayesian approach to obtained posterior pdf of states. With the help of Bayes’ theorem we write 
\begin{equation*}
    p(X_k|y_{1:k}) = \dfrac{p(y_k|X_k)p(X_k|y_{1:k-1})}{p(y_k|y_{1:k-1})}.
\end{equation*}
The predicted measurement is given as
\begin{equation}
    p(y_k|y_{1:k-1}) = \mathcal{N}(y_k;\hat{y}_{k|k-1},P_{yy,k|k-1}),
\end{equation}
where \begin{equation} \label{eqy}
    \hat{y}_{k|k-1} = \int h(X_k)\mathcal{N}(X_k;\hat{X}_{k|k-1},\bar{P}_{k|k-1})dX_k.
\end{equation}
The innovation covariance matrix, which is formulated using modified predicted error covariance and modified measurement noise covariance, is given as
\begin{equation}\label{eqpy} \begin{split}
    P_{yy,k|k-1} = & \int h(X_k)h(X_k)^\top \mathcal{N}(X_k;\hat{X}_{k|k-1},\bar{P}_{k|k-1})dX_k - \\ 
    & \hat{y}_{k|k-1}\hat{y}_{k|k-1}^\top+\bar{R}_k.
\end{split}\end{equation}
Similarly, the cross-covariance matrix is
\begin{equation}\label{eqpxy} \begin{split}
    P_{Xy,k|k-1} = & \int X_kh(X_k)^\top \mathcal{N}(X_k;\hat{X}_{k|k-1},\bar{P}_{k|k-1})dX_k- \\
    & \hat{X}_{k|k-1}\hat{y}_{k|k-1}^\top.
\end{split}\end{equation}
The posterior state estimate is given by
\begin{equation}\label{eqxkk}
    \hat{X}_{k|k} =  \hat{X}_{k|k-1}+K_k(y_k-\hat{y}_{k|k-1}),
\end{equation} 
where 
\begin{equation} \label{eqnkk}
     K_k =  P_{Xy,k|k-1}P_{yy,k|k-1}^{-1}. 
 \end{equation}
 The expression for posterior error covariance  is 
 \begin{equation}\label{eqpkk}
       P_{k|k} =  \bar{P}_{k|k-1}-K_kP_{yy,k|k-1}K_k^\top.
 \end{equation}  
 To calculate the posterior estimate, correntropy matrix, $\mathit{\Pi}_k$, needs to be calculated and for that posterior estimate, $\hat{X}_{k|k}$ should be known. To circumvent the problem, a fixed point iteration is required which iteratively calculates $\hat{X}_{k|k}$ for a fixed $k$ till the state estimation error between two consecutive iterations becomes significantly small.  

 Like the Gaussian filters, maximum correntropy filters approximate the prior and posterior pdfs as Gaussian and represent with mean and covariance. 
 The first attempt at maximum correntropy filter for a non-linear system is presented in MC-EKF \cite{liu2019linear}. It is implemented using the Jacobian matrix, and a regression model is used to develop a fixed point iterative algorithm to obtain the posterior state estimate \cite{liu2016extended}. 
The MC-EKF suffers from a few drawbacks such as smoothness requirement of system and measurement function, large estimation error and lack of convergence, which are inherited from the ordinary EKF \cite{bhaumik2019nonlinear}.

To overcome the disadvantages of MC-EKF, a few filtering methods such as MC-UKF \cite{liu2016maximum}, MC-CKF \cite{liu2018maximum},  and MC-NSKF \cite{urooj20222d} are developed. The said filters evaluate the integrals mentioned in (\ref{eqx}), (\ref{eqpk}), (\ref{eqy}), (\ref{eqpy}), and (\ref{eqpxy}) using a few carefully chosen deterministic sample points and their weights. In this paper, we propose to use polynomial chaos expansion (PCE) to linearize nonlinear equations \cite{wiener1938homogeneous} while implementing the maximum correntropy criteria. 
\section{ Maximum Correntropy Polynomial Chaos Kalman Filter}
\subsection{Hermite polynomial chaos expansion}
Let us assume that at any instant of time, state variable $X$ follows a Gaussian distribution with mean $\hat{X}$ and covariance $P=\mathcal{S}\mathcal{S}^\top$. This pdf can be expressed with the help of collocation points located at $(\hat{X}+\mathcal{S}\xi)$, where $\xi_j$ follows the standard normal distribution and dimension of $\xi$ depends on the order of the Hermite polynomial. 
Considering a $p^{th}$ order Hermite PCE, we can expand a function $f(.)$ as
\begin{equation} \begin{split}
    f(\hat{X}+\mathcal{S}\xi) =& a_0+\sum_{m_1=1}^{n}a_{m_1}H_1(\xi_{m_1})+\sum_{m_1=1}^{n}\sum_{m_2=1}^{m_1}a_{m_1m_2}\\
    & H_2(\xi_{m_1},\xi_{m_2})+\ldots+\sum_{m_1=1}^n\ldots \sum_{m_p=1}^{m_{p-1}} a_{m_1 i_2\ldots m_p}\\
    & H_p(\xi_{m_1},\xi_{m_2},\ldots,\xi_{m_p}),
\end{split}\end{equation}
where $H_p(\xi_{m_1}\ldots \xi_{m_p})$ is $p^{th}$ order the Hermite polynomial, $a_{m_1},a_{m_1m_2}\ldots$ are deterministic coefficients of the Hermite polynomial which are to be calculated. The expression for Hermite polynomial of order $p$ is given as,
   \begin{equation}\begin{split}
       H_{p} (\xi_{m_1},\xi_{m_2}\ldots \xi_{m_n}) =&  (-1)^p \exp(\dfrac{1}{2}\xi\xi^\top)\dfrac{\partial^n}{\partial \xi_{m_1}\partial \xi_{m_2} \ldots \partial \xi_{m_n}} \\
       & \exp(\dfrac{1}{2}\xi^\top \xi).
   \end{split}\end{equation}
It is interesting to note that the Hermite polynomial is orthogonal, which means that in inner product sense $\mathbb{E}[H_iH_j] =0$ for $i\neq j$. Higher accuracy in approximation can be achieved using the higher-order Hermite PCE. However, the improvement in accuracy for $p>2 \ or\ 3$ is negligible, with a tremendous increase in computational burden.
   The second-order ($p=2$) Hermite PCE of the process function is
   \begin{equation} \label{eqnpce2ndorder}\begin{split}
       f(\hat{X}+\mathcal{S}\xi) \approx & \ a_0+\sum_{m_1=1}^{n}a_{m_1}\xi_{m_1}+\sum_{m_1=1}^{n}a_{m_1 m_1}(\xi_{m_1}^2-1) +\\ &\sum_{m_1=1}^{n}\sum_{m_2=1}^{m_1}a_{m_1m_2}\xi_{m_1}\xi_{m_2}.
   \end{split}  \end{equation}
   
A system with $n$ states and $p^{th}$ order Hermite polynomial series contains $\binom{n+p}{p}$ number of coefficients. So, for a moderately high dimensional system, computational burden is considerably high. To reduce the computational burden, throughout our work we consider a second order Hermite polynomial \emph{i.e} $p=2$,  and last term in the above equation (`cross terms' as per \cite{kumar2023polynomial}) is ignored. Under such truncation the Hermite polynomial contains $2n+1$ number of coefficients, which is a good bargain in terms of computational burden with intact system accuracy. Under the above approximation \eqref{eqnpce2ndorder} becomes
\begin{equation*} \begin{split}
       f(\hat{X}+\mathcal{S}\xi) \approx & \ a_0+\sum_{m_1=1}^{n}a_{m_1}H_1(\xi_{m_1})+\sum_{m_1=1}^{n}a_{m_1 m_1}H_2(\xi_{m_1},\\
       & \xi_{m_2})= a_0+A_{n\times 2n}\hat{H}, \\
      \end{split} \end{equation*}
      or,
      \begin{equation}    \label{eqcoeff}        
      f(\hat{X}+\mathcal{S}\xi) = [  a_0 \ \ A]_{n\times (2n+1)} [I \ \ H  ]^\top_{1 \times (2n+1)}= \hat{A}\hat{H},
   \end{equation}
   where $\hat{A}_{n\times (2n+1)}=\begin{bmatrix} a_0 & A \end{bmatrix}^{\top}$ is the coefficient matrix for process function \cite{ren2015probabilistic}. The Hermite polynomial matrix is $H_{(2n+1)\times (2n)}=\begin{bmatrix}  & H_1(\xi_{1})  & \ldots & H_1(\xi_{n}) & H_2(\xi_{1}) & \ldots & H_2(\xi_{n})\end{bmatrix}$. Similarly, the measurement function can be approximated as
\begin{equation}\label{eqpcmeasurement}
            h(\hat{X}+\mathcal{S}\xi) =  b_o+BH
             =  \hat{B}\hat{H},
    \end{equation}
where $\hat{B}_{m\times (2n+1)}=\begin{bmatrix} b_0 & B \end{bmatrix}^{\top}$ is the coefficient matrix for the measurement function. The approximated process and measurement model can be written as
\begin{equation}\label{eqpcmodel}\begin{split}
        X_k =& \ a_0+AH+\mu_k,\\
        y_k = & \ b_0+BH+r_k.
    \end{split}\end{equation}   
The procedure to obtain the unknown coefficients matrices $\hat{A}$ and $\hat{B}$ includes the following steps. Firstly, the collocation points ($\xi$) are generated, which are used to get the Hermite polynomial matrix. Then CPs are spread using error covariance matrix and functions are approximated on those points. Finally, the coefficient matrices are obtained by solving the equation (\ref{eqcoeff}) and \eqref{eqpcmeasurement}.
 \subsection{Generation of deterministic sample points}
As we consider a second order PCE, we have to find out the roots of $3^{rd}$ order Hermite polynomial which are ($0, -\sqrt{3},\sqrt{3}$). The collocation points are the intersection points of a $n$ dimensional hypersphere and the coordinate axes. So for a $n$ dimensional system we receive $2n+1$ number of collocation points. Mathematically, all the collocation points are represented by a matrix $\xi= \begin{bmatrix} diag(-\sqrt{3})_{n\times n} & 0_{n\times 1} & diag(\sqrt{3})_{n \times n} \end{bmatrix}_{n \times (2n+1)}$, where each column represents a single collocation point in $n$ dimensional real space. So $m^{th}$ CP points is expressed with $\xi_{m_{(n \times 1)}}=\begin{bmatrix} \xi_{1m} & \xi_{2m} & \ldots & \xi_{nm} \end{bmatrix}^\top$. 
The expression of Hermite polynomial matrix ($\hat{H}$) is
 \begin{equation*} \hat{H}^\top= \begin{bmatrix} 1 & \cdots & 1 & \ldots & 1 \\
         H_1(\xi_{1,1}) & \cdots & H_1(\xi_{1,j}) & \cdots & H_1(\xi_{1,2n+1})\\
          H_1(\xi_{2,1}) & \cdots & H_1(\xi_{2,j}) & \cdots &  H_1(\xi_{2,2n+1})\\
         \vdots & \vdots & \vdots  & \vdots & \vdots \\
         H_1(\xi_{n,1}) & \cdots  & H_1(\xi_{n,j}) & \cdots & H_1(\xi_{n,2n+1}) \\
         H_2(\xi_{1,1}) & \cdots &  H_2(\xi_{1,j}) & \cdots & H_2(\xi_{1,2n+1})\\
         H_2(\xi_{2,1}) & \cdots & H_2(\xi_{2,j}) & \cdots & H_2(\xi_{2,2n+1})\\
         \vdots & \vdots & \vdots & \vdots & \vdots \\
         H_2(\xi_{n,1}) & \cdots & H_2(\xi_{n,j}) & \cdots & H_2(\xi_{n,2n+1})
     \end{bmatrix}.\end{equation*}

\subsection{Maximum correntropy polynomial chaos Kalman filter}
\textit{1) Prior estimation:} The sample points ($\mathcal{X}_j$) are generated around the posterior mean $\hat{X}_{k-1|k-1}$ using the CPs as given below
    \begin{equation} \label{eqpckf1}
        \mathcal{X}_{k|k-1,j} = \hat{X}_{k-1|k-1}+\mathcal{S}_{k-1|k-1}\xi_j,\ j = 1,\ldots, 2n+1, 
    \end{equation}
    where  $\mathcal{S}_{k-1|k-1}$ is the square-root of the posterior error covariance matrix, \emph{i.e.} $P_{k-1|k-1}=\mathcal{S}_{k-1|k-1}\mathcal{S}_{k-1|k-1}^\top$.
 The sample points $\mathcal{X}_j$ are propagated through the process function to yield a set of new points,
  \begin{equation} \label{eqpckf2}
      \mathcal{X}_{k|k-1,j} = f(\mathcal{X}_{k-1|k-1,j}).
  \end{equation}
    These transformed samples are concatenated to form 
    $\mathcal{X}_{k|k-1} = \begin{bmatrix} \mathcal{X}_{k|k-1,1} & \mathcal{X}_{k|k-1,2} & \cdots & \mathcal{X}_{k|k-1,2n+1}
        \end{bmatrix}^\top.$
The Hermite coefficient matrix for the process function is obtained as
    \begin{equation} \label{eqhermite}
            \hat{A} = (\hat{H}^{-1}\mathcal{X}_{k|k-1})^\top.
        \end{equation}
  The prior mean is obtained using equation   \eqref{eqpcmodel} in \eqref{eqnprior} as given below,
    \begin{equation} \label{pcxprior} \begin{split}
        \hat{X}_{k|k-1} =  & \  \mathbb{E}[(a_0+AH+\mu_k)|y_{1:k-1}]\\
        = & \ a_0=\hat{A}_1^\top,
    \end{split}\end{equation}
    where $\hat{A}_{1}$ is the first row of coefficient matrix, $\hat{A}$. The prior error covariance matrix using equations \eqref{eqpk} and \eqref{eqpcmodel} is calculated as
    \begin{equation}\label{pcpprior}\begin{split}
        P_{k|k-1} = & \mathbb{E}[(AH+\mu_k)(AH+\mu_k)^\top|y_{1:k-1}]\\
        = & AA^\top +Q.
    \end{split}\end{equation}
  \textit{2) Posterior estimation:}  The transformed samples set using measurement function is given as
   \begin{equation}
       \mathcal{Y}_{k|k-1,j}=h(\hat{X}_{k|k-1}+\mathcal{S}_{k|k-1}\xi_j), \ j=1,\ldots,2n+1.
   \end{equation}
The coefficient matrix for the measurement function using equation (\ref{eqpcmeasurement}) is given as, 
    \begin{equation}
        \hat{B} = (\hat{H}^{-1}\mathcal{Y}_{k|k-1})^\top,
    \end{equation}
    where $ \mathcal{Y}_{k|k-1} = \begin{bmatrix} \mathcal{Y}_{k|k-1,1} & \mathcal{Y}_{k|k-1,2} & \cdots & \mathcal{Y}_{k|k-1,2n+1}
        \end{bmatrix}^\top.$
  The predicted measurement is
  \begin{equation}\label{pcy}\begin{split}
      \hat{y}_{k|k-1} = & \mathbb{E}[(b_0+BH+r_k)|y_{1:k-1}]\\
      = & b_0= \hat{B}_{1}^\top,
 \end{split} \end{equation}
 where $\hat{B}_1$ is the first row of the coefficient matrix for measurement function, $\hat{B}$. 
 
To calculate the posterior state estimate $\hat{X}_{k|k}$, we need to know the value of correntropy matrix $\mathit{\Pi}_{k}$, the modified prior error covariance $\bar{P}_{k|k-1}$, and modified measurement noise covariance $\bar{R}_{k}$, which are the function of error vector $e_{k,l}$ mentioned in \eqref{eqekj}. The calculation of error vector demands the availability of true value of state $X_k$. 
As we don't know the truth $X_k$, we use fixed point iteration (FPI) technique which is started by assigning the prior state estimate as the truth value, \emph{i.e.} $\hat{X}_{k}^{0}=\hat{X}_{k|k-1}$ \cite{chen2017maximum,chen2019minimum,liu2016extended} and calculate $e_{k,l}^{i}$, $\mathit{\Pi}_{k}^{i}$, $\bar{P}_{k|k-1}^{i}$, $\bar{R}_{k}^{i}$, and $\hat{X}_{k|k}^{i+1}$ using \eqref{eqekj}, \eqref{eq29}, \eqref{eqp}, \eqref{eqr} and \eqref{eqxkk}, respectively. Here, we use the superscript $i$ to denote iteration step at any time step $k$. 
As we are starting with the approximate truth, we need to run the whole set of above mentioned equations iteratively till the consecutive posterior estimate are close enough.
  
 The iterated innovation error covariance matrix is calculated as \begin{equation}\label{pcpyy}\begin{split}
     P_{yy,k|k-1}^i = & \mathbb{E}[(y_k-y_{k|k-1})(y_k-y_{k|k-1})^\top|y_{1:k-1}]\\
     = & \mathbb{E}[(BH+r_k)(BH+r_k)^\top|y_{1:k-1}]\\
     = & BB^\top+\bar{R}_k^{i}.
 \end{split}\end{equation}
The cross-covariance matrix is calculated before performing the FPI by numerically approximating the integration \eqref{eqpxy} using the coefficients as given below
\begin{equation} \label{pcpxy}\begin{split}
    P_{Xy,k|k-1} =& \mathbb{E}[(X_k-\hat{X}_{k|k-1})(y_k-\hat{y}_{k|k-1})|y_{1:k-1}]\\
    = & AB^\top.
\end{split}\end{equation}
The expression for Kalman gain at  $i^{th}$ iteration is
\begin{equation} \label{eqnkknew}
     K_k^{i} =  P_{Xy,k|k-1}(P_{yy,k|k-1}^{i})^{-1}. 
 \end{equation}
 Finally the posterior state estimate at each iterative step is obtained using (\ref{eqxkk}), given as
       \begin{equation}\label{eqxkknew}
    \hat{X}_{k|k}^{i+1} =  \hat{X}_{k|k-1}+K_k^{i}(y_k-\hat{y}_{k|k-1}).
\end{equation} 
This iteration will continue till the relative error $\frac{||\hat{X}_{k|k}^{i+1} - \hat{X}_{k|k}^{i}||}{||\hat{X}_{k|k}^{i}||} \leq \epsilon$, a user defined parameter.
Once the fixed point iteration is completed, the posterior error covariance is calculated using (\ref{eqpkk}).

The FPI algorithm can be initialized using the prior estimate \cite{chen2017maximum,chen2019minimum,liu2016extended}.
In such way of initialization, although the initial state error at $0^{th}$ iteration, ($e_{k,l}^0$) becomes zero, it does not led to any computational error. However, in this work, we calculate posterior estimate of state using the traditional PCKF and initialize the FPI algorithm with it. As in the later method, the state is initialized with an approximated posterior estimate, it is expected to converge faster. The FPI algorithm is provided in Algorithm \ref{fpialgorithm}.
\begin{algorithm}
\caption{Fixed point iteration (FPI)}
\label{fpialgorithm}
   $[\hat{X}_{k|k}]: = FPI [\hat{X}_{k|k-1}, y_k]$
    \begin{itemize}
        \item Calculate $\hat{X}^{PCKF}_{k|k}$ with PCKF and initialize $\hat{X}_{k|k}^i= \hat{X}^{PCKF}_{k|k}$.
        \item \textbf{for} $i=1:i_{max}$
        \begin{itemize}
            \item Calculate $e_{k,l}^{i}$, $\Pi_{p,k}^{i}$ and $\Pi_{R,k}^{i}$  using (\ref{eqekj}), (\ref{eq29}).
            \item Evaluate $\bar{P}_{k|k-1}^i$ and $\bar{R}_k^i$ using (\ref{eqp}) and (\ref{eqr}).
            \item Calculate $P_{yy,k|k-1}^{i}$ using (\ref{pcpyy}).
            \item Compute $K_k^{i}$ and $\hat{X}_{k|k}^{i+1}$ using (\ref{eqnkknew}) and (\ref{eqxkknew}). 
            \item \textbf{if} $\frac{||\hat{X}_{k|k}^{i+1} - \hat{X}_{k|k}^{i}||}{||\hat{X}_{k|k}^{i}||} \leq \epsilon$
            \begin{itemize}
		\item[-] $\hat{X}_{k|k} = \hat{X}_{k|k}^{i+1}$.
		\item [-] break
		\end{itemize}
          \textbf{else} $i = i + 1$.
        \end{itemize}
          \item \textbf{end for}       
    \end{itemize}
\end{algorithm}
\subsection{Square-root implementation of PCKF} 
The calculation of modified error covariance matrix, modified measurement noise covariance matrix and sample points generations around the posterior mean (see the Eqns. (\ref{eqp}), (\ref{eqr}) and (\ref{eqpckf1})) demands the calculation of the square-root of the error covariance $P_{k|k-1}$, measurement noise covariance $R_k$ and posterior error covariance $P_{k|k}$, respectively. It is usually done using the Cholesky decomposition. Sometimes the Cholesky decomposition can lead to negative definiteness of the matrix due to round-off errors during computation \cite{bhaumik2013cubature,van2001square}. These limitations are well documented and it is tackled using the QR decomposition technique \cite{arasaratnam2008square,bhaumik2014square}, eliminating the need to explicitly square root the covariance matrix.

The error covariance matrix $P_{k|k-1}\in \mathbb{R}^{n \times n}$ is factorized as $P_{k|k-1}= UU^\top$. The QR decomposition of matrix $U \in \mathbb{R}^{n \times p}$, $n\geq p$, is $U = \mathcal{R}_1^\top \mathcal{Q}_1^\top$, where $\mathcal{Q}_1\in \mathbb{R}^{p \times p}$ is orthogonal and $\mathcal{R}_1 \in  \mathbb{R}^{p \times n}$ is upper triangular matrix. So, 
\begin{equation*} 
     P_{k|k-1} =  UU^\top =  \mathcal{R}_1^\top \mathcal{Q}_1^\top \mathcal{Q}_1 \mathcal{R}_1= \mathfrak{P}^\top \mathfrak{P}=\mathcal{S}_{k|k-1}\mathcal{S}_{k|k-1}^\top.
     \end{equation*} 
     Similarly,
 \begin{equation*}
  P_{k|k} = MM^\top =  \mathcal{R}_2^\top \mathcal{Q}_2^\top \mathcal{Q}_2 \mathcal{R}_2= \mathfrak{P'}^\top \mathfrak{P'}=\mathcal{S}_{k|k}\mathcal{S}_{k|k}^\top,\\
\end{equation*}
\begin{equation*}
     R_{k} = OO^\top = \mathcal{R}_3^\top \mathcal{Q}_3^\top \mathcal{Q}_3 \mathcal{R}_3= \mathfrak{R}^\top \mathfrak{R} = \mathcal{S}_{R,k}\mathcal{S}_{R,k}^\top,
\end{equation*}
where $\mathfrak{P}\in \mathbb{R}^{n \times n}$, $\mathfrak{P'}\in \mathbb{R}^{n \times n}$ and $\mathfrak{R}\in \mathbb{R}^{m \times m}$ are the upper triangular part of $\mathcal{R}_1$, $\mathcal{R}_2$ and $\mathcal{R}_3$ matrices, respectively.
The generalized representation of square-root of covariance matrix using QR decomposition is provided below, where \textit{qr} and \textit{uptri} represents the QR decomposition and upper triangular part of matrix, respectively.
\begin{equation} \label{eqprqr}\begin{split}
    \mathcal{R}_{1,k|k-1} =& \ qr[P_{k|k-1}],\\
    \mathcal{R}_{2,k} = & \ qr[P_{k|k}],\\
    \mathcal{R}_{3,k} = & \ qr[R_{k}].
\end{split}\end{equation}
The square root of the covariance matrix are the upper triangular matrices obtained from the QR decomposition as given below
\begin{equation} \label{eqsqrpkk-1}
    \mathcal{S}_{k|k-1}^\top=\mathfrak{P}_{k|k-1}= \ uptri\{\mathcal{R}_{1,k|k-1}\},\end{equation}
\begin{equation}\label{eqsqrpkk}\mathcal{S}_{k|k}^\top=\mathfrak{P}_{k|k}= \ uptri\{\mathcal{R}_{2,k|k-1}\},\end{equation}
    \begin{equation}\label{eqsqrr}\mathcal{S}_{R,k}^\top=\mathfrak{R}_{k} = \ uptri\{\mathcal{R}_{3,k}\}.
\end{equation}
 The filtering steps for the square root maximum correntropy polynomial chaos Kalman filter (SR-MC-PCKF) are summarized in Algorithm \ref{MC-SR-PCKF}.
\begin{algorithm} 
\caption{ SR-MC-PCKF}
\label{MC-SR-PCKF}
\begin{itemize}
    \item Set initial values of $X_{0|0}$ and $P_{0|0}$.
    \item \textbf{for} $k=1:k_{max}$
          \begin{itemize}
              \item Calculate $\mathcal{S}_{k-1|k-1}$ using \eqref{eqsqrpkk}.
              \item Calculate $\hat{X}_{k|k-1}$ and $P_{k|k-1}$ using (\ref{pcxprior}) and (\ref{pcpprior}).
              \item Calculate $\mathcal{S}_{k|k-1}$ and $\mathcal{S}_{R,k}$ using \eqref{eqsqrpkk-1} and \eqref{eqsqrr}.
              \item Calculate $\hat{y}_{k|k-1}$ using (\ref{pcy}).
              \item Calculate $P_{Xy,k|k-1}$ using (\ref{pcpxy}).
              \item Calculate $\hat{X}_{k|k} = FPI[\hat{X}_{k|k-1}, y_k]$.
              \item Calculate $P_{k|k}$ using  (\ref{eqpkk}).
          \end{itemize}
          \item \textbf{end for}
\end{itemize}  
\end{algorithm}
\section{Computational Complexity}
 The computational complexity of an algorithm is measured in terms of floating point operations (flop) counts. The addition of two matrix $A\in \mathbb{R}^{n \times m}$ and $B\in \mathbb{R}^{n \times m}$ requires $nm$ flops. The multiplication of matrix $A\in \mathbb{R}^{n \times m}$ and $B\in \mathbb{R}^{m \times n}$ requires $n^2(2m-1)$ flops. The QR decomposition and inverse operation on matrix $A\in \mathbb{R}^{n \times n}$ requires $(4/3)n^3$ and $n^3$ flops, respectively \cite{arasaratnam2008square}.
The computational complexity of EKF, square root PCKF, and square root SPKF are 
\begin{equation*} 
     \mathcal{C}_{EKF} =  4n^3+m^3+6n^2m+6m^2n-nm+n+2m,
     \end{equation*}
     \begin{equation*} \begin{split}
     \mathcal{C}_{SPKF} = & \frac{8}{3}n^3+m^3-\frac{2}{3}N_p^3+n^2(9N_p+1)+2m^2N_p+\\
                                   &2N_p^2  (n+m)+2n^2m+4m^2n+mn(2N_p-\\
                                   &1)+ 2N_p(2n+m)+n+2m+2N_p,\end{split}\end{equation*}
\begin{equation*} \begin{split}   \mathcal{C}_{PCKF} = & \dfrac{8}{3}n^3+m^3+2N_p^2(n+m)+N_p(6n^2-n- 
                         m+\\
                         & 2m^2+2nm)-2n^2-2m^2-3nm+4nm^2+\\
                         & m+2n^2m,
\end{split} \end{equation*}
respectively, where $N_p$ is the number of sigma points. The UKF, CKF and NSKF uses $(2n+1)$, $2n$ and $(4n+1)$ respective number of sigma points. The $N_p$ in PCKF represents the number of collocation points, which is $(2n+1)$ in our case.

For MC filters, FPI needs to be executed. The flops count for FPI algorithms for different filters are as follows
 \begin{equation*} \begin{split}
     \mathcal{C}_{FPI,EKF} =& \mathcal{C}_{EKF}+\dfrac{16}{3}(n^3+m^3)+4n^2+2m^2+4(n^2m\\
     & +m^2n)+3nm+3m,\end{split} \end{equation*} \begin{equation*} \begin{split}
     \mathcal{C}_{FPI,SPKF} =& \mathcal{C}_{SPKF}+\frac{16}{3}(n^3+m^3)+4n^2+2m^2(N_p+\\
     & 1)+ 2mnN_p+2m^2n+n+4m,\end{split} \end{equation*}\begin{equation*} \begin{split}
     \mathcal{C}_{FPI,PCKF}  = & \mathcal{C}_{PCKF} + \dfrac{16}{3}n^3+\dfrac{19}{3}m^3+4n^2+2m^2(1+\\
     & N_p+n)+nm+4m+n.
 \end{split} \end{equation*}
Assuming that the average number of required fixed point iterations is $T$, the flops count of MC-EKF, MC-SPKF and MC-PCKF are,
 \begin{equation*} \begin{split}
     \mathcal{C}_{MC-EKF} = & \mathcal{C}_{EKF}+T\mathcal{C}_{FPI,EKF}-\big[ m^3+4nm(n+m)\\
                            & + 3nm+m \big],\end{split} \end{equation*}
\begin{equation*}  \begin{split}  \mathcal{C}_{MC-SPKF} = & \mathcal{C}_{SPKF}+T\mathcal{C}_{FPI,SPKF}-\big[ m^3+2m^2N_p\\
    & +2nm(N_p-1)+2nm(m+1) +m \big],\end{split} \end{equation*} \begin{equation*} \begin{split}
     \text{and} \ \mathcal{C}_{MC-PCKF} = & \mathcal{C}_{PCKF}+T\mathcal{C}_{FPI,PCKF}-\big[ m^3+ 2m^2\\
    & (N_p+m)+nm+m \big].
\end{split} \end{equation*}
From the above equations we see that flops count of all the three filters are comparable. Flops count are little more when we use maximum correntropy filtering.
\section{Simulation Results}
In order to assess the effectiveness of the proposed MC-PCKF filter, an underwater real-life navigation problem is solved using the proposed method and its performance is compared with the existing estimators for a simulated underwater trajectory. The square-root version of the estimators are implemented to avoid any unexpected stop of the implemented filters during execution.  
The INS mechanization equations are highly non-linear, and the EKF handles the non-linearity by taking the first-order Taylor series expansion, which introduces approximation error. Among the deterministic sample point filters, Gauss-Hermite filter (GHF) suffers from the `curse of dimensionality' problem due to a large number of support points requirement. So, we discard implementing the GHF and the EKF. The state estimation is performed using the process model and measurement model as given in section \ref{probemformulation}. 
   In simulation, we have considered the latitude and longitude measurements from APS to be available up to $200^{th}$ time steps and follow the measurement model II. The APS becomes unavailable after $200^{th}$ time step, so after that measurement model I is followed.
    \begin{figure}[ht]
\centerline{\includegraphics[scale =0.5]{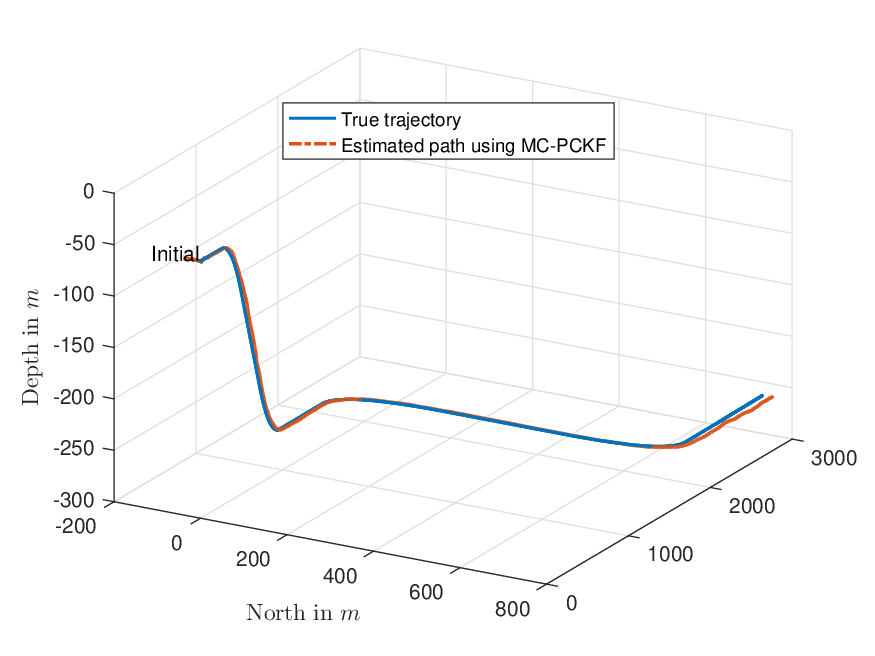}}
    	\caption{True path along with estimated path using MC-PCKF.}
    	\label{tracking}
    	\end{figure} 
\subsection{Engagement Scenario}
The vehicle is initially submerged in water at a depth of $50m$, at stationary condition. The initial latitude and longitude of the vehicle are $18.946^o$N and $72.854^o$E, respectively. An acceleration input was applied along the east direction, resulting the body to attain a velocity of $v^n=\begin{bmatrix}0 & 10 & 0 \end{bmatrix}^\top knots$ at the end of $100 \ sec$. Following this the vehicle performs a series of manoeuvres during the entire motion period of $900 \ sec$. The vehicle dynamics are according to the parameters outlined in Table \ref{table3},  which mentions the acceleration input in $m/sec^2$ and the angular rate in $deg/sec$. All manoeuvres performed by the vehicle are within the ability of the older class of submarine and the manoeuvrability limits are taken from \cite{joubert2004some}. The roll rate, pitch rate and yaw rate are taken as $0.4^o/sec,0.4^o/sec\ \text{and}\ 1.8^o/sec$ respectively. The true trajectory of the vehicle is shown in Fig. \ref{tracking}. The position of GPS intelligent buoys on the sea surface in terms of $(L,l, Z)$ are $GIB1=\begin{bmatrix} 18.9461^oN & 72.8541^oE & 0 \end{bmatrix}$ and $GIB2=\begin{bmatrix} 18.9459^oN & 72.8539^oE & 0 \end{bmatrix}$. The sampling rate of all the sensors is taken as 1 Hz. 

 The process noise, $Q$ is taken as
$Q  = diag(  0_{1 \times 3},  \ (5 \times 10^{-5} g)^2I_{ 1 \times 3}, \ (\frac{0.02^o}{\sqrt{1hr}})^2 I_{1 \times 3})$. 
We follow \cite{liu2018maximum} to define the measurement noise of latitude ($L$), longitude ($l$), and height as $Z$, as 
$r_{L} , r_{l}\sim 0.9\mathcal{N}( 0,(0.0898^o)^2 ) + 0.1\mathcal{N}( 0,(0.898^o)^2 )$, $ r_{Z} \sim 0.9  \mathcal{N}( 0, (1)^2 ) + 0.1  \mathcal{N}( 0, (10)^2 )$, respectively. Following \cite{liu2018maximum} we define the noise for velocity measurement in NED direction as 
$r_{v^N} \sim 0.9\mathcal{N}( 0, (0.1)^2 )+0.1 \mathcal{N}( 0, (1)^2 )$,  $r_{v^E} \sim0.9\mathcal{N}( 0, (0.1)^2 )+0.1 \mathcal{N}( 0, (1)^2 )$, $r_{v^D} \sim 0.9 \mathcal{N}( 0, (0.1)^2 )+0.1 \mathcal{N}( 0, (1)^2 )$, respectively. The measurement noise for roll, pitch and yaw are taken as 
$r_{\phi}, r_{\theta}, r_{\psi} \sim 0.9\mathcal{N}( 0, (0.5^o)^2 )+0.1\mathcal{N}( 0, (1^o)^2 )$.

To have an unbiased comparison among the estimators, they are given the same initialization, $\hat{X}_{0|0} = \begin{bmatrix} 18.944^oN & 72.853^oE & -25 & 0 & 0 & 0 &  0 & 0 & 0 \end{bmatrix}^\top$, initial error covariance, $P_{0|0}=diag(  (0.898^o)^2,   (0.898^o)^2,  \ (10)^2,  \  (2)^2, \   (2)^2,   \ (2)^2,  \  (1^o)^2, \\ \ (1^o)^2, \ (5^o)^2  )$.
The true path for the underwater vehicle along with the estimated path using the proposed estimator, MC-PCKF is shown in Fig. \ref{tracking}. The estimated path divergence from the true trajectory due to APS blackout is visually justified from the plot.

\begin{table*}[htbp] 
	 	\caption{Underwater vehicle manoeuvre in the simulation}
	 	\begin{center}
	 		\begin{tabular} {p{1cm} p{1.5cm} p{3cm} p{1cm} p{1cm} p{2cm} p{1cm} p{1cm} p{1cm}}
	 			\hline 
	 			Stage & Period (sec)& Vehicle manoeuvre & \multicolumn{6}{c}{Parameters} \\
	 			& & & $a^N$ &  $a^E$ & $a^D$ & $\dot{\phi}$ & $\dot{\theta}$ & $\dot{\psi}$\\
                  \hline
                  1 & 0-100  & Constant acc. & $0$ &  $\dfrac{1}{20}$ & $-g$ & $0$ & $0$ & $0$\\
                  2 & 101-150  & Pitch down & $0$ &  $0.018$ & $-(g+0.04)$ & $0$ & $\dfrac{2}{5}$ & $0$\\
                  3 & 151-200  & Uniform linear motion & $0$ &  $0$ & $-g$ & $0$ & $0$ & $0$\\
                  4 & 201-250  & Pitch up & $0$ &  $0.018$ & $-(g-0.04)$ & $0$ & $-\dfrac{2}{5}$ & $0$\\
                  5 & 251-350  & Uniform linear motion & $0$ & $0$ & $-g$ & $0$ & $0$ & $0$\\
                  6 & 351-355  & Roll motion & $0$ & $0$ & $-g$ & $\dfrac{2}{5}$ & $0$ & $0$\\
                  7 & 356-450  & Turn right & $0.053$ &  $-\dfrac{1}{20}$ & $-g$ & $0$ & $0$ & $\dfrac{19}{20}$\\
                  8 & 451-455  & Roll motion &  $0$ &  $0$ & $-g$ & $-\dfrac{2}{5}$ & $0$ & $0$\\
                  9 & 456-500 & Uniform linear motion &  $0$ &  $0$ & $-g$ & $0$ & $0$ & $0$\\
                  10 & 501-505 & Roll motion &  $0$ &  $0$ & $-g$ & $-\dfrac{2}{5}$ & $0$ & $0$\\
                  11 & 506-600 & Turn left & $-0.053$ &  $\dfrac{1}{20}$ & $-g$ & $0$ & $0$ & $\dfrac{19}{20}$\\
                  12 & 601-605 & Roll angle turn & $0$ &  $0$ & $-g$ & $\dfrac{2}{5}$ & $0$ & $0$\\
                  13 & 606-650 & uniform linear vel. & $0$ & $0$ & $-g$ & $0$ & $0$ & $0$\\
                  14 & 651-700 & Constant decel. & $0$ &  $-\dfrac{1}{20}$ & $-g$ & $0$ & $0$ & $0$\\
                  15 & 701-900 & Uniform linear motion & $0$ &  $0$ & $-g$ & $0$ & $0$ & $0$\\
        \hline
	 		\end{tabular}
	 		\label{table3}
	 	\end{center}
	 \end{table*}
	\subsection{Performance Comparison}
\begin{figure}[ht]
\centerline{\includegraphics[scale =0.48]{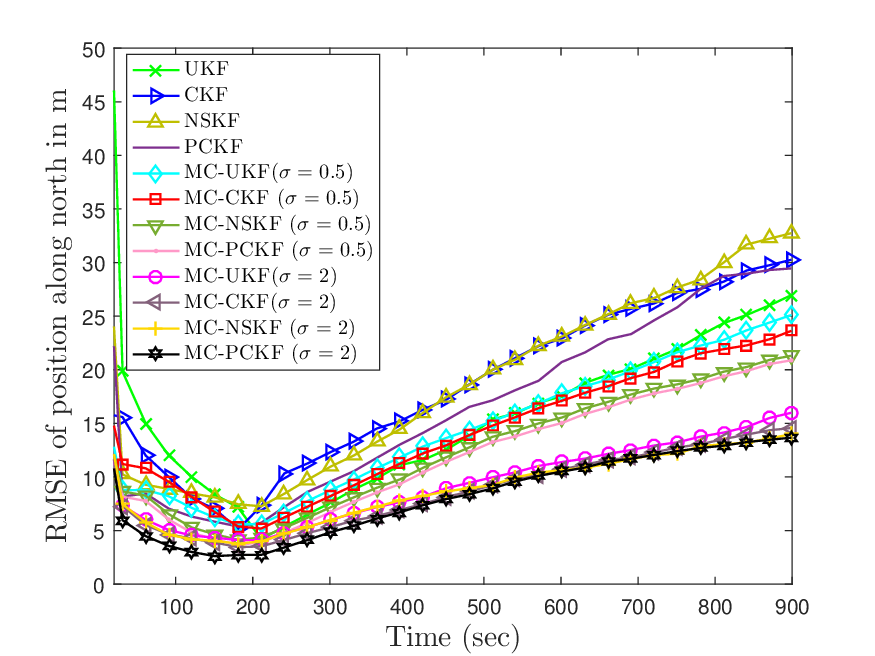}}
    	\caption{The RMSE of position along north direction.}
    	\label{rmse_lat}
    	\end{figure}

\begin{figure}[ht]
    	\centerline{\includegraphics[scale =0.48]{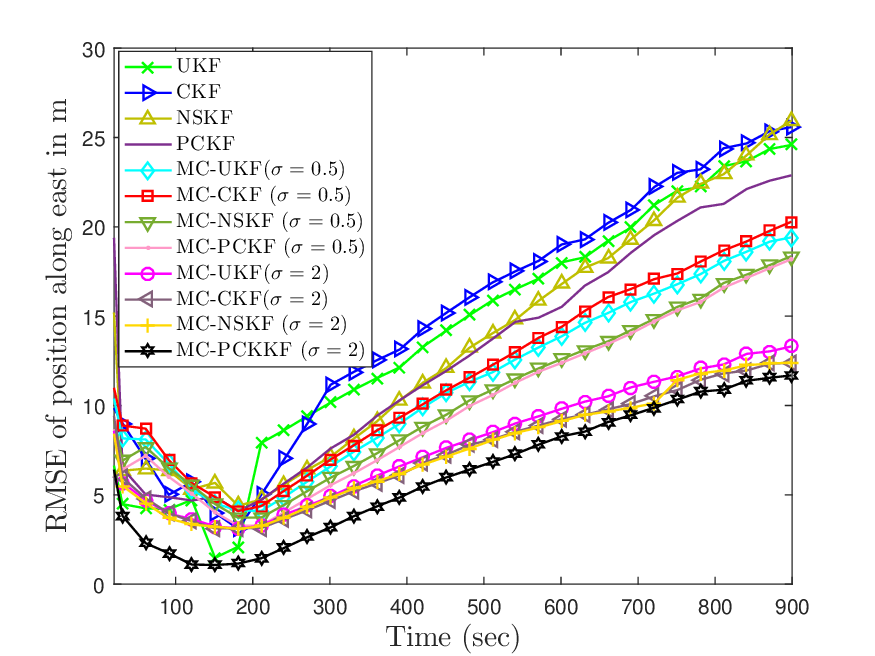}}
    	\caption{The RMSE of position along east direction.}
    	\label{rmse_long}
    \end{figure}

\begin{figure}[ht]
	\centerline{\includegraphics[scale =0.48]{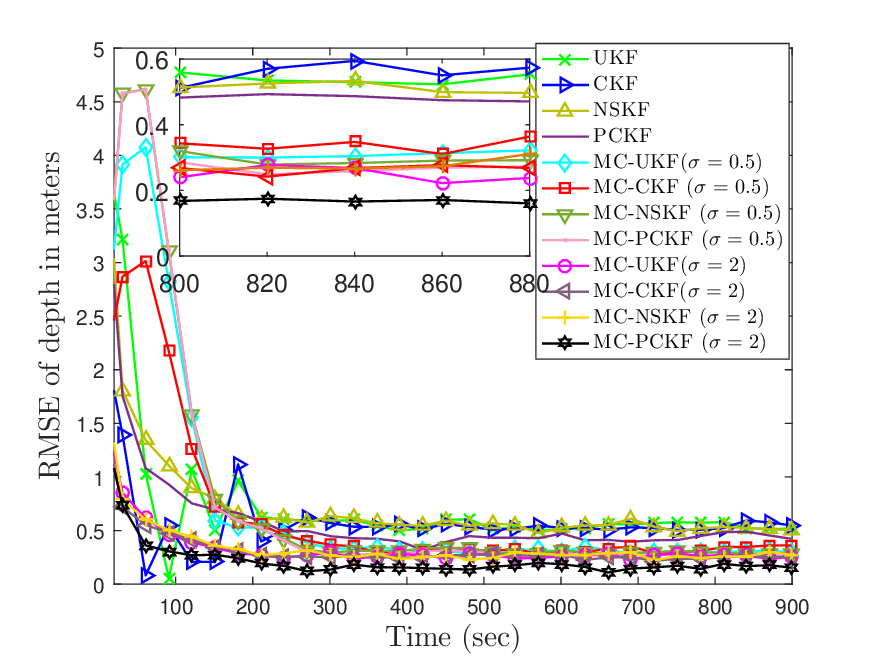}}
	\caption{The RMSE of depth.}
	\label{rmse_h}
\end{figure}

\begin{figure}[ht]
	\centerline{\includegraphics[scale =0.48]{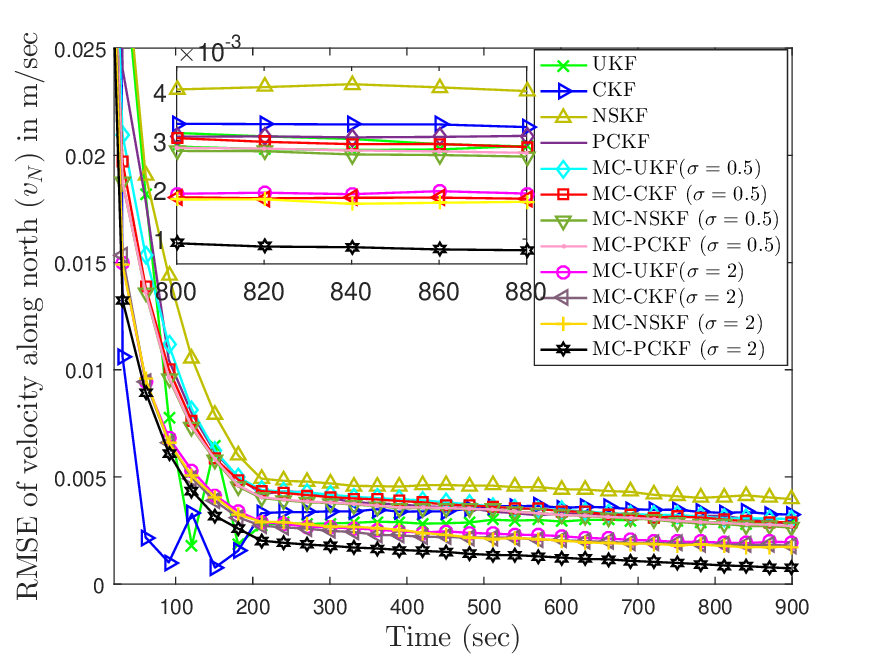}}
	\caption{The RMSE of velocity along north direction.}
	\label{rmse_vN}
\end{figure}

\begin{figure}[ht]
	\centerline{\includegraphics[scale =0.48]{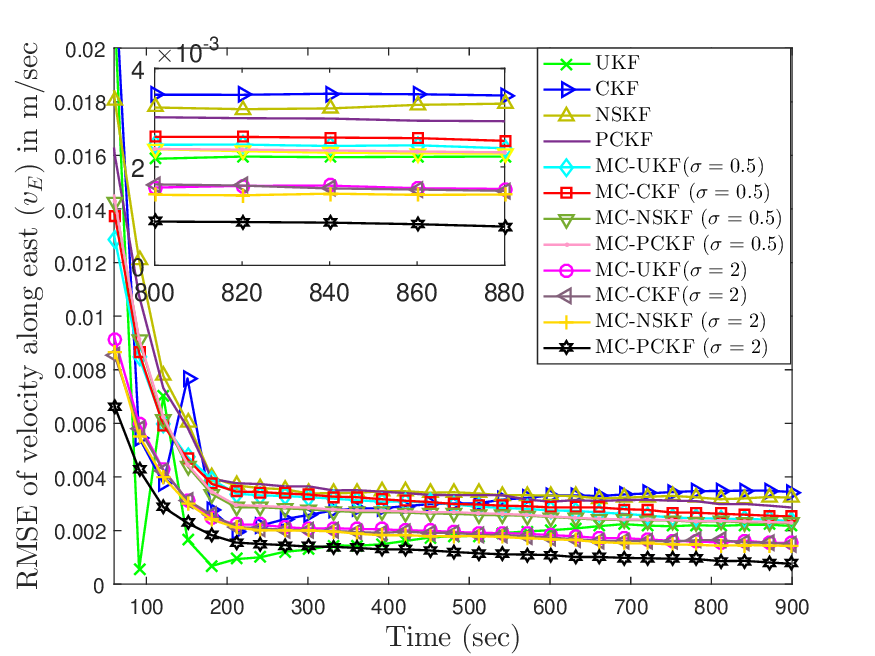}}
	\caption{The RMSE of velocity along east direction.}
	\label{rmse_vE}
\end{figure}

\begin{figure}[ht]
	\centerline{\includegraphics[scale =0.48]{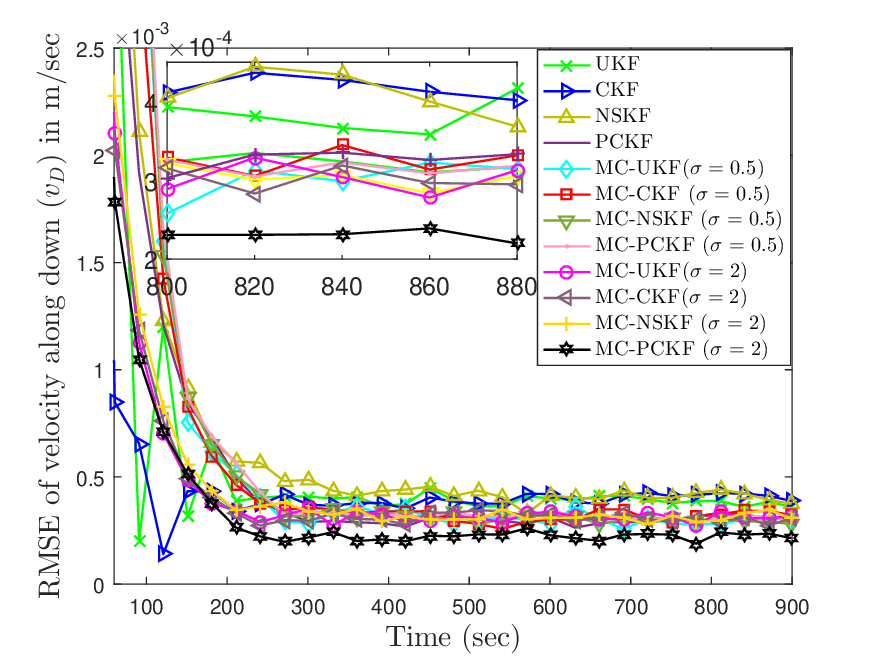}}
	\caption{The RMSE of velocity along down direction.}
	\label{rmse_vD}
\end{figure}

\begin{figure}[ht]
	\centerline{\includegraphics[scale =0.48]{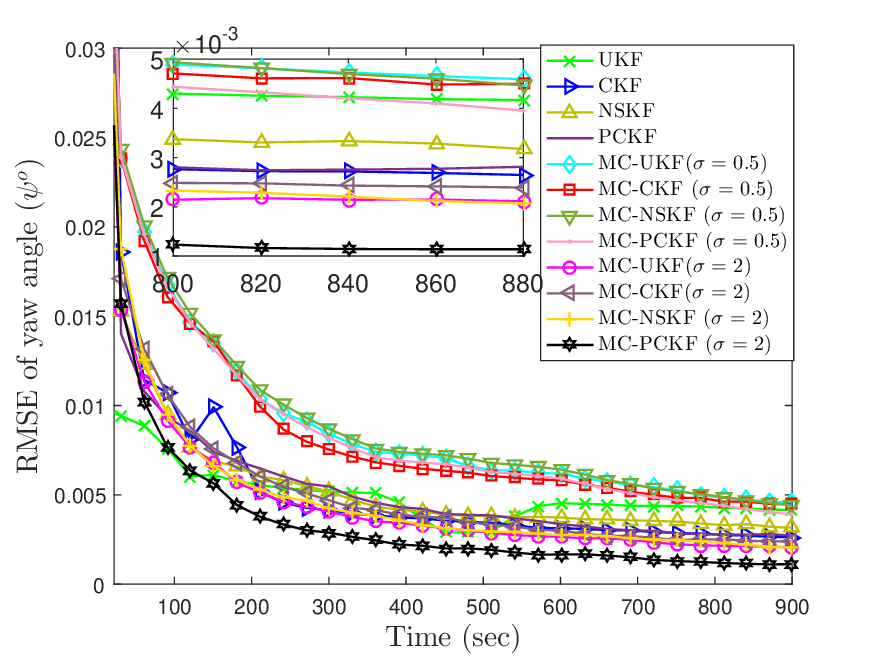}}
	\caption{The RMSE of yaw angle $\theta$.}
	\label{rmse_yaw}
\end{figure}

The performance of the various estimators is compared on benchmarks such as RMSE, averaged RMSE (ARMSE) and the execution time. RMSE of any state $x_k$ at any time instant $k$ is calculated over MC ensemble using the formula, $RMSE_k = \frac{1}{MC}\sum_{i=1}^{MC}\sqrt{(x_{k,i}-\hat{x}_{k|k,i})^2} $. The performance of the proposed MC-PCKF filter is compared in presence of non-Gaussian noises in terms of RMSE plot. Please note that all the filters are implemented using QR decomposition which is popularly known as square root filtering, but for simplicity we avoid mentioning it repeatedly. So, from now onward if we mention a filter's name it means the square root version of it. 
RMSE is calculated out of 100 independent Monte Carlo runs, and the total simulation time of a single run is 900 sec.
The RMSE of position estimated along north, east and down (NED) directions are plotted in Fig. \ref{rmse_lat}, \ref{rmse_long}, and \ref{rmse_h}, respectively. 
From the figures, we see that MC filters with $\sigma=2$ provide lower RMSE compared to their mean square error (MSE) counterparts. 
Please note that, the latitude and longitude measurements from the APS are available only up to $200 \ sec$ and after that the APS signal is lost and system becomes unobservable. As a result, the RMSE values for latitude and longitude start increasing after that time, which can be seen from Fig. \ref{rmse_lat} and \ref{rmse_long}. Interesting to note that among all the filters the rate of increase of RMSE is lowest in MC-PCKF with $\sigma=2$.
The RMSE plots for velocity estimate along NED direction are given Fig. \ref{rmse_vN}, \ref{rmse_vE}, and \ref{rmse_vD}. It can be observed that the lowest RMSE is obtained using MC-PCKF with $\sigma=2$. 
The RMSE plot for yaw angle estimation is given in Fig. \ref{rmse_yaw} and similar plots are obtained for other orientation angles such as roll and pitch which are not shown here. There is no divergence in RMSE during velocity and attitude estimation.

The averaged RMSE values obtained using different filters for non-Gaussian noises is listed in Table \ref{SumGau}. 
It can be observed that ARMSEs of the maximum correntropy filters are less than the simple sigma point filters and MC-PCKF is the lowest among them.  
The computational demands of all the filters are calculated in terms of flops count and relative execution time and they are listed in Table \ref{computation}. From the table, it can be observed that the computational demand of MC-PCKF is comparable with other sigma point MC filters such as MC-CKF and MC-UKF. The MC-NSKF has the highest computational demand as it requires approximately twice as much as sigma points compared to other filters.
		\begin{table*}[htbp] 
		\caption{ARMSEs of position, velocity and orientation for various filters}
		\begin{center}
			\begin{tabular}{ |p{3cm}| p{1.1cm}p{1.1cm} p{1.1cm}| p{1.1cm} p{1.1cm} p{1.1cm} |p{1.1cm} p{1.1cm} p{1.1cm}| }
					\hline 
					Filter & $x^N$ & $x^E$ & $x^D$ & $v^N$  &  $v^E$  & $v^D$ &  roll angle ($\phi^o$) & pitch angle ($\theta^o$) & yaw angle ($\psi^o$) \\
				\hline
				UKF & 18.14 & 14.813 & 0.756 & 0.01 & 0.0033 & 0.0018  & 0.086 & 0.0887 & 0.1277\\
				CKF & 17.299 & 14.0938 & 0.742 & 0.009 & 0.0029 & 0.00175  & 0.085 & 0.0844 & 0.122\\
                NSKF & 17.177 & 13.23 & 0.695 & 0.0094 & 0.0032 & 0.00173  & 0.083 & 0.0749 & 0.1331\\
                PCKF & 16.675 & 13.027 & 0.661 & 0.0088 & 0.0026 & 0.0016  & 0.08 & 0.07 & 0.1030\\
                MC-UKF($\sigma=0.5$) & 13.202 & 10.0136 & 0.728 & 0.0075 & 0.0022 & 0.00099 & 0.072 & 0.05281 & 0.1099\\
				MC-CKF($\sigma=0.5$) & 12.881 & 10.582 & 0.739 & 0.0078 & 0.0023 & 0.00091 & 0.081 & 0.0553 & 0.098\\
                MC-NSKF($\sigma=0.5$) & 11.739 & 9.680 & 0.784 & 0.0076 & 0.002 & 0.001 & 0.076 & 0.064 & 0.094\\
                 MC-PCKF($\sigma=0.5$) & 10.615 & 9.8633 & 0.6652 & 0.0071 & 0.0018 & 0.0009 & 0.066 & 0.048 & 0.083\\
				MC-UKF($\sigma=2$) & 7.8871 & 7.2135 & 0.435 & 0.0033 & 0.0014 & 0.000677 & 0.027 & 0.0264 & 0.0484 \\
				MC-CKF($\sigma=2$) & 7.5461 & 6.9947 & 0.417 & 0.0031 &0.0012 & 0.000566 & 0.029 & 0.0275 & 0.0497 \\ 
				MC-NSKF($\sigma=2$) & 6.67 & 6.91 & 0.406 & 0.0032 & 0.0011 & 0.00056 & 0.0244 & 0.0245 & 0.045\\
                MC-PCKF($\sigma=2$) & 6.264 & 6.09 & 0.341 & 0.00284 & 0.001 & 0.000499 & 0.0207 & 0.0194 & 0.0324\\
				\hline
			\end{tabular}
			\label{SumGau}
		\end{center}
	\end{table*}
	\begin{table*}[htbp]
		\caption{Comparison in terms of flops count and relative computational time}
		\begin{center}
			\begin{tabular}{p{2cm}p{1cm}p{1cm}p{1cm}p{1cm}p{1cm}p{1.2cm}p{1.25cm}p{1.2cm}p{1.2cm}p{1.25cm}}\\
				 \hline 
                Filters  & PCKF & CKF & UKF & NSKF  & MC-PCKF & MC-CKF & MC-UKF & MC-NSKF\\ \hline \hline
                Flops count  & 34533 & 34812 & 36568 & 63622  & 77364 & 77193 & 80706 & 134810 \\ \hline
                Relative execution time  & 1 & 1.008 & 1.06 & 1.84  & 2.24 & 2.24 & 2.34 & 3.90 \\
				\hline
			\end{tabular}
			\label{computation}
		\end{center}
	\end{table*}


\section{Conclusion}
The paper proposed the maximum correntropy polynomial chaos Kalman filter (MC-PCKF) for estimation of states in presence of non-Gaussian noise. Additionally, a sensor fusion method for underwater vehicle navigation is developed, utilizing measurements from a carefully selected set of auxiliary sensors. Emphasis is given to ensure not to use any active sensor which makes the vehicle undetectable by enemy tracking systems. The proposed filter was successfully applied for underwater navigation to estimate position and attitude of the vehicle. Comparative analysis with existing deterministic sample point filters revealed that the proposed filter exhibits better performance in presence of heavily tailed non-Gaussian noise. Also, the computational load of the proposed algorithm is comparable with the existing filters. Overall, the findings of this study highlight the efficacy and applicability of the proposed MC-PCKF algorithm in sensor fusion in presence of non-Gaussian noise, emphasizing its potential to enhance underwater navigation system.

\section*{Declaration of competing interest}
The authors declare that they have no known competing financial interests or personal relationships that could have appeared to influence the work reported in this paper.
\section*{Acknowledgements}
The second author is financially supported by scholarships from the Prime Minister's Research Fellows (PMRF) Scheme (PMRF id: 2702440).


\begin{thebibliography}{10}

\bibitem{zhuang2023multi}
Y.~Zhuang, X.~Sun, Y.~Li, J.~Huai, L.~Hua, X.~Yang, X.~Cao, P.~Zhang, Y.~Cao, L.~Qi \emph{et~al.}, ``Multi-sensor integrated navigation/positioning systems using data fusion: From analytics-based to learning-based approaches,'' \emph{Information Fusion}, vol.~95, pp. 62--90, 2023.

\bibitem{kumar2021recent}
M.~Kumar and S.~Mondal, ``Recent developments on target tracking problems: A review,'' \emph{Ocean Engineering}, vol. 236, p. 109558, 2021.

\bibitem{zhang2023autonomous}
B.~Zhang, D.~Ji, S.~Liu, X.~Zhu, and W.~Xu, ``Autonomous underwater vehicle navigation: A review,'' \emph{Ocean Engineering}, p. 113861, 2023.

\bibitem{sahoo2019advancements}
A.~Sahoo, S.~K. Dwivedy, and P.~Robi, ``Advancements in the field of autonomous underwater vehicle,'' \emph{Ocean Engineering}, vol. 181, pp. 145--160, 2019.

\bibitem{paull2013auv}
L.~Paull, S.~Saeedi, M.~Seto, and H.~Li, ``A{UV} navigation and localization: A review,'' \emph{IEEE Journal of Oceanic Engineering}, vol.~39, no.~1, pp. 131--149, 2013.

\bibitem{leonard2016autonomous}
J.~J. Leonard and A.~Bahr, ``Autonomous underwater vehicle navigation,'' \emph{Springer handbook of {O}cean {E}ngineering}, pp. 341--358, 2016.

\bibitem{brokloff1997dead}
N.~A. Brokloff, ``Dead reckoning with an {ADCP} and current extrapolation,'' in \emph{Oceans' 97. MTS/IEEE Conference Proceedings}, vol.~2.\hskip 1em plus 0.5em minus 0.4em\relax IEEE, 1997, pp. 1411--vol.

\bibitem{groves2015principles}
P.~D. Groves, ``Principles of {GNSS}, inertial, and multisensor integrated navigation systems,'' \emph{IEEE Aerospace and Electronic Systems Magazine}, vol.~30, no.~2, pp. 26--27, 2015.

\bibitem{kinsey2006survey}
J.~C. Kinsey, R.~M. Eustice, and L.~L. Whitcomb, ``A survey of underwater vehicle navigation: Recent advances and new challenges,'' in \emph{IFAC Conference of Manoeuvering and Control of Marine Craft}, vol.~88.\hskip 1em plus 0.5em minus 0.4em\relax Lisbon, 2006, pp. 1--12.

\bibitem{zhang2020geomagnetic}
J.~Zhang, T.~Zhang, H.-S. Shin, J.~Wang, and C.~Zhang, ``Geomagnetic gradient-assisted evolutionary algorithm for long-range underwater navigation,'' \emph{IEEE Transactions on Instrumentation and Measurement}, vol.~70, pp. 1--12, 2020.

\bibitem{wang2016particle}
B.~Wang, L.~Yu, Z.~Deng, and M.~Fu, ``A particle filter-based matching algorithm with gravity sample vector for underwater gravity aided navigation,'' \emph{IEEE/ASME Transactions on Mechatronics}, vol.~21, no.~3, pp. 1399--1408, 2016.

\bibitem{duecker2020towards}
D.~A. Duecker, N.~Bauschmann, T.~Hansen, E.~Kreuzer, and R.~Seifried, ``Towards micro robot hydrobatics: Vision-based guidance, navigation, and control for agile underwater vehicles in confined environments,'' in \emph{2020 IEEE/RSJ International Conference on Intelligent Robots and Systems (IROS)}.\hskip 1em plus 0.5em minus 0.4em\relax IEEE, 2020, pp. 1819--1826.

\bibitem{sobreira2019map}
H.~Sobreira, C.~M. Costa, I.~Sousa, L.~Rocha, J.~Lima, P.~Farias, P.~Costa, and A.~P. Moreira, ``Map-matching algorithms for robot self-localization: a comparison between perfect match, iterative closest point and normal distributions transform,'' \emph{Journal of Intelligent \& Robotic Systems}, vol.~93, pp. 533--546, 2019.

\bibitem{miller2010autonomous}
P.~A. Miller, J.~A. Farrell, Y.~Zhao, and V.~Djapic, ``Autonomous underwater vehicle navigation,'' \emph{IEEE Journal of Oceanic Engineering}, vol.~35, no.~3, pp. 663--678, 2010.

\bibitem{davari2016asynchronous}
N.~Davari and A.~Gholami, ``An asynchronous adaptive direct {K}alman filter algorithm to improve underwater navigation system performance,'' \emph{IEEE Sensors Journal}, vol.~17, no.~4, pp. 1061--1068, 2016.

\bibitem{allotta2016unscented}
B.~Allotta, A.~Caiti, L.~Chisci, R.~Costanzi, F.~Di~Corato, C.~Fantacci, D.~Fenucci, E.~Meli, and A.~Ridolfi, ``An unscented {K}alman filter based navigation algorithm for autonomous underwater vehicles,'' \emph{Mechatronics}, vol.~39, pp. 185--195, 2016.

\bibitem{urooj20222d}
A.~Urooj, A.~Dak, B.~Ristic, and R.~Radhakrishnan, ``2{D} and 3{D} angles-only target tracking based on maximum correntropy {K}alman filters,'' \emph{Sensors}, vol.~22, no.~15, p. 5625, 2022.

\bibitem{wan2000unscented}
E.~A. Wan and R.~Van Der~Merwe, ``The unscented {K}alman filter for nonlinear estimation,'' in \emph{Proceedings of the IEEE 2000 Adaptive Systems for Signal Processing, Communications, and Control Symposium}.\hskip 1em plus 0.5em minus 0.4em\relax IEEE, 2000, pp. 153--158.

\bibitem{arasaratnam2009cubature}
I.~Arasaratnam and S.~Haykin, ``Cubature {K}alman filters,'' \emph{IEEE Transactions on Automatic Control}, vol.~54, no.~6, pp. 1254--1269, 2009.

\bibitem{radhakrishnan2018new}
R.~Radhakrishnan, A.~Yadav, P.~Date, and S.~Bhaumik, ``A new method for generating sigma points and weights for nonlinear filtering,'' \emph{IEEE Control Systems Letters}, vol.~2, no.~3, pp. 519--524, 2018.

\bibitem{bhaumik2019nonlinear}
S.~Bhaumik and P.~Date, \emph{Nonlinear estimation: methods and applications with deterministic sample points}.\hskip 1em plus 0.5em minus 0.4em\relax CRC Press, 2019.

\bibitem{kumar2023polynomial}
K.~Kumar, R.~K. Tiwari, S.~Bhaumik, and P.~Date, ``Polynomial chaos {K}alman filter for target tracking applications,'' \emph{IET Radar, Sonar \& Navigation}, vol.~17, no.~2, pp. 247--260, 2023.

\bibitem{liu2016maximum}
X.~Liu, H.~Qu, J.~Zhao, P.~Yue, and M.~Wang, ``Maximum correntropy unscented {K}alman filter for spacecraft relative state estimation,'' \emph{Sensors}, vol.~16, no.~9, p. 1530, 2016.

\bibitem{liu2018maximum}
X.~Liu, H.~Qu, J.~Zhao, and P.~Yue, ``Maximum correntropy square-root cubature {K}alman filter with application to {SINS}/{GPS} integrated systems,'' \emph{ISA Transactions}, vol.~80, pp. 195--202, 2018.

\bibitem{xu2018novel}
Y.~Xu, L.~Mili, and J.~Zhao, ``A novel polynomial-chaos-based {K}alman filter,'' \emph{IEEE Signal Processing Letters}, vol.~26, no.~1, pp. 9--13, 2018.

\bibitem{chen2017maximum}
B.~Chen, X.~Liu, H.~Zhao, and J.~C. Principe, ``Maximum correntropy {K}alman filter,'' \emph{Automatica}, vol.~76, pp. 70--77, 2017.

\bibitem{van2001square}
R.~Van Der~Merwe and E.~A. Wan, ``The square-root unscented {K}alman filter for state and parameter-estimation,'' in \emph{2001 IEEE International Conference on Acoustics, Speech, and Signal Processing. Proceedings}, vol.~6.\hskip 1em plus 0.5em minus 0.4em\relax IEEE, 2001, pp. 3461--3464.

\bibitem{mu2021practical}
X.~Mu, B.~He, S.~Wu, X.~Zhang, Y.~Song, and T.~Yan, ``A practical ins/gps/dvl/ps integrated navigation algorithm and its application on autonomous underwater vehicle,'' \emph{Applied Ocean Research}, vol. 106, p. 102441, 2021.

\bibitem{hemingway2018perspectives}
E.~G. Hemingway and O.~M. O’Reilly, ``Perspectives on {E}uler angle singularities, gimbal lock, and the orthogonality of applied forces and applied moments,'' \emph{Multibody System Dynamics}, vol.~44, pp. 31--56, 2018.

\bibitem{crassidis2006sigma}
J.~L. Crassidis, ``Sigma-point kalman filtering for integrated {GPS} and inertial navigation,'' \emph{IEEE Transactions on Aerospace and Electronic Systems}, vol.~42, no.~2, pp. 750--756, 2006.

\bibitem{farrell2008aided}
J.~Farrell, \emph{Aided navigation: GPS with high rate sensors}.\hskip 1em plus 0.5em minus 0.4em\relax McGraw-Hill, Inc., 2008.

\bibitem{klein2015observability}
I.~Klein and R.~Diamant, ``Observability analysis of {DVL}/{PS} aided {INS} for a maneuvering {AUV},'' \emph{Sensors}, vol.~15, no.~10, pp. 26\,818--26\,837, 2015.

\bibitem{healey1998online}
A.~J. Healey, E.~An, and D.~Marco, ``Online compensation of heading sensor bias for low cost {AUV}s,'' in \emph{Proceedings of the 1998 Workshop on Autonomous Underwater Vehicles}.\hskip 1em plus 0.5em minus 0.4em\relax IEEE, 1998, pp. 35--42.

\bibitem{shaukat2021multi}
N.~Shaukat, A.~Ali, M.~Javed~Iqbal, M.~Moinuddin, and P.~Otero, ``Multi-sensor fusion for underwater vehicle localization by augmentation of {RBF} neural network and error-state {K}alman filter,'' \emph{Sensors}, vol.~21, no.~4, p. 1149, 2021.

\bibitem{marco2001command}
D.~B. Marco and A.~J. Healey, ``Command, control, and navigation experimental results with the {NPS} {ARIES} {AUV},'' \emph{IEEE Journal of Oceanic Engineering}, vol.~26, no.~4, pp. 466--476, 2001.

\bibitem{kebkal2017auv}
K.~G. Kebkal and A.~Mashoshin, ``{AUV} acoustic positioning methods,'' \emph{Gyroscopy and Navigation}, vol.~8, no.~1, pp. 80--89, 2017.

\bibitem{kinsey2007situ}
J.~C. Kinsey and L.~L. Whitcomb, ``In situ alignment calibration of attitude and doppler sensors for precision underwater vehicle navigation: Theory and experiment,'' \emph{IEEE Journal of Oceanic Engineering}, vol.~32, no.~2, pp. 286--299, 2007.

\bibitem{lee2007simulation}
P.-M. Lee, B.-H. Jun, K.~Kim, J.~Lee, T.~Aoki, and T.~Hyakudome, ``Simulation of an inertial acoustic navigation system with range aiding for an autonomous underwater vehicle,'' \emph{IEEE Journal of Oceanic Engineering}, vol.~32, no.~2, pp. 327--345, 2007.

\bibitem{liu2007correntropy}
W.~Liu, P.~P. Pokharel, and J.~C. Principe, ``Correntropy: Properties and applications in non-{G}aussian signal processing,'' \emph{IEEE Transactions on Signal Processing}, vol.~55, no.~11, pp. 5286--5298, 2007.

\bibitem{liu2016extended}
X.~Liu, H.~Qu, J.~Zhao, and B.~Chen, ``Extended {K}alman filter under maximum correntropy criterion,'' in \emph{2016 International Joint Conference on Neural Networks (IJCNN)}.\hskip 1em plus 0.5em minus 0.4em\relax IEEE, 2016, pp. 1733--1737.

\bibitem{liu2019linear}
X.~Liu, Z.~Ren, H.~Lyu, Z.~Jiang, P.~Ren, and B.~Chen, ``Linear and nonlinear regression-based maximum correntropy extended {K}alman filtering,'' \emph{IEEE Transactions on Systems, Man, and Cybernetics: Systems}, vol.~51, no.~5, pp. 3093--3102, 2019.

\bibitem{singh2010closed}
A.~Singh and J.~C. Principe, ``A closed form recursive solution for maximum correntropy training,'' in \emph{2010 IEEE International Conference on Acoustics, Speech and Signal Processing}.\hskip 1em plus 0.5em minus 0.4em\relax IEEE, 2010, pp. 2070--2073.

\bibitem{zhao2022robust}
H.~Zhao, B.~Tian, and B.~Chen, ``Robust stable iterated unscented {K}alman filter based on maximum correntropy criterion,'' \emph{Automatica}, vol. 142, p. 110410, 2022.

\bibitem{wiener1938homogeneous}
N.~Wiener, ``The homogeneous chaos,'' \emph{American Journal of Mathematics}, vol.~60, no.~4, pp. 897--936, 1938.

\bibitem{ren2015probabilistic}
Z.~Ren, W.~Li, R.~Billinton, and W.~Yan, ``Probabilistic power flow analysis based on the stochastic response surface method,'' \emph{IEEE Transactions on Power Systems}, vol.~31, no.~3, pp. 2307--2315, 2015.

\bibitem{chen2019minimum}
B.~Chen, L.~Dang, Y.~Gu, N.~Zheng, and J.~C. Pr{\'\i}ncipe, ``Minimum error entropy {K}alman filter,'' \emph{IEEE Transactions on Systems, Man, and Cybernetics: Systems}, vol.~51, no.~9, pp. 5819--5829, 2019.

\bibitem{bhaumik2013cubature}
S.~Bhaumik \emph{et~al.}, ``Cubature quadrature {K}alman filter,'' \emph{IET Signal Processing}, vol.~7, no.~7, pp. 533--541, 2013.

\bibitem{arasaratnam2008square}
I.~Arasaratnam and S.~Haykin, ``Square-root quadrature {K}alman filtering,'' \emph{IEEE Transactions on Signal Processing}, vol.~56, no.~6, pp. 2589--2593, 2008.

\bibitem{bhaumik2014square}
S.~Bhaumik and Swati, ``Square-root cubature-quadrature {K}alman filter,'' \emph{Asian Journal of Control}, vol.~16, no.~2, pp. 617--622, 2014.

\bibitem{joubert2004some}
P.~Joubert, ``Some aspects of submarine design part 1. hydrodynamics,'' Defence Science and Technology Organisation Victoria (Australia), Tech. Rep., 2004.

\end{thebibliography}
\end{document}